%% file: 00_main.tex
\documentclass{article}

\usepackage{fullpage}

\usepackage{tikz}
\usepackage{hyperref}
\usepackage{amsmath}
\usepackage{amsthm}
\usepackage{booktabs}
\usepackage{graphicx}
\usepackage{subcaption}
\usepackage{float}
\usepackage[normalem]{ulem}
\theoremstyle{definition}
\usepackage{multirow}
\usepackage{array} 
\newcolumntype{P}[1]{>{\raggedright\arraybackslash}p{#1}}

\definecolor{mediumgreen}{RGB}{60, 179, 113}
\definecolor{darkgreen}{rgb}{0.0, 0.5, 0.0}

\definecolor{comment-red}{rgb}{0.8,0,0}
\definecolor{lightgray}{gray}{0.7}

\newcommand\shortsection[1]{\vspace{6pt}{\noindent\bf #1.}}

\usepackage[backend=biber,sorting=nyt,sortcites,maxbibnames=99]{biblatex}
\usepackage{biblatex}
\title{What did Elon change? A comprehensive analysis of Grokipedia}
\author{Harold Triedman\thanks{HT is a part-time contractor at Wikimedia Enterprise. This work was conducted independently from Wikimedia Enterprise and the Wikimedia Foundation, and the perspectives expressed here are solely his own.} ~and Alexios Mantzarlis \\
Cornell Tech \\
\texttt{\{hjt36,amantzarlis\}@cornell.edu}}
\date{}

\begin{document}

\maketitle

\begin{abstract}
    Elon Musk released Grokipedia on 27 October 2025 to provide an alternative to Wikipedia, the crowdsourced online encyclopedia. In this paper, we provide the first comprehensive analysis of Grokipedia and compare it to a dump of Wikipedia, with a focus on article similarity and citation practices. Although Grokipedia articles are much longer than their corresponding English Wikipedia articles, we find that much of Grokipedia's content (including both articles with and without Creative Commons licenses) is highly derivative of Wikipedia. Nevertheless, citation practices between the sites differ greatly, with Grokipedia citing many more sources deemed ``generally unreliable'' or ``blacklisted'' by the English Wikipedia community and low quality by external scholars, including dozens of citations to sites like Stormfront and Infowars. We then analyze article subsets: one about elected officials, one about controversial topics, and one random subset for which we derive article quality and topic. We find that the elected official and controversial article subsets showed less similarity between their Wikipedia version and Grokipedia version than other pages. The random subset illustrates that Grokipedia focused rewriting the highest quality articles on Wikipedia, with a bias towards biographies, politics, society, and history. Finally, we publicly release our nearly-full scrape of Grokipedia, as well as embeddings of the entire Grokipedia corpus. 
\end{abstract}

\input{01_intro}
\input{02_methods}
\input{03_findings}
\input{05_conclusion}

\section*{Acknowledgments}
Many thanks to Mykola Trokhymovych for his help in framing interesting analyses would be interesting, and for suggestions about how to scrape Grokipedia.

\printbibliography
\clearpage
\appendix
\input{99_appendix}

\end{document}

%% file: 01_intro.tex
\section{Introduction}

On 27 October 2025, Elon Musk launched Grokipedia as an AI-powered alternative to Wikipedia. The tech billionaire had previously attacked the crowdsourced encyclopedia as an ``extension of legacy media propaganda''\footnote{\url{https://x.com/elonmusk/status/1881752812276891674}}; he vowed that Grokipedia's goal would be ``the truth, the whole truth and nothing but the truth.''\footnote{\url{https://x.com/elonmusk/status/1983009037156315440}} 

At launch, Grokipedia was composed of 885,279 entries. Media reports and spot checks of individual entries found that Musk's encyclopedia appeared to be more opinionated than English Wikipedia, adding editorial slants that appeared to align with Musk's own political views \cite{indicatorBriefingGrokipediaApplies, rogersElonMusksGrokipedia, wongWhatElonMusks2025}. A recent review of the 1,800 most-edited articles on English Wikipedia that also have entries on Grokipedia conducted by Taha Yasseri found that the latter appears to be highly derivative of the former but ``privileges fluency and narrative over attribution'' \cite{yasseriHowSimilarAre2025}.

Our paper builds on this work by conducting the first full-scale comparison of the two services. Between 28 and 30 October 2025, we were able to successfully scrape nearly the entire Grokipedia corpus (883,858 articles, or 99.8\%). By reviewing the text and citation similarity of this corpus with its Wikipedia equivalent, we seek to provide more data to determine whether Musk's product is in fact a ``synthetic derivative'' \cite{yasseriHowSimilarAre2025}, an ideological project \cite{wongWhatElonMusks2025}, or something else altogether.

The initial release of Grokipedia is a tale of two licenses. A majority of Grokipedia's content (56\% of scraped articles) is basically derivative of English Wikipedia, porting over the latter's Creative Commons (CC) Attribution. Grokipedia's CC-licensed articles have, on average, a 90\% similarity to their corresponding Wikipedia article. The remainder of Grokipedia's content is (according to xAI's November 2025 Terms of Service\footnote{\url{https://archive.ph/58VTS}}) licensed in the same way as the outputs of the Grok chatbot. These non-CC-licensed articles are notably less similar to their corresponding English Wikipedia articles (77\% similarity). 

At a high level, the two services base their encyclopedic articles frequently on the same sources, with 57 shared domains among the top 100 most-cited sources across the two encyclopedias. However source composition differs: Grokipedia leans more heavily on academic sources, government sources, and user-generated content compared to Wikipedia's relative over-reliance on news websites. 

Grokipedia articles are much longer and more verbose than Wikipedia, and cite twice as many sources. This comes at the expense of source quality. Using source categorizations for domains determined by English Wikipedia editors and domain reliability scores from \cite{linHighLevelCorrespondence2023}, we find that Grokipedia relies on a lower share of reliable sources and a higher share of unreliable sources than English Wikipedia.

The dichotomy between CC- and non-CC-licensed content is also visible in source citations. Non-CC-licensed articles on Grokipedia are 3.2 times more likely than the same articles on Wikipedia to contain a citation that the English Wikipedia community has deemed ``generally unreliable'' and 13 times more likely to contain a ``blacklisted'' source. It is noteworthy, even if the number of citations is trivial as an overall share of sources, that Grokipedia includes 42 citations for Nazi website Stormfront and 34 for conspiracy website InfoWars, compared to none for both on English Wikipedia.

In conjuction with this paper, we release a full, structured scrape of the content of Grokipedia, as well as embeddings of Grokipedia chunks. These datasets are available for download on HuggingFace\footnote{\url{https://huggingface.co/datasets/htriedman/grokipedia-v0.1-dump}}, and our analysis code is available on GitHub\footnote{\url{https://github.com/htried/wiki-grok-comparison}}. 

%% file: 02_methods.tex
\section{Methodology}
\label{sec:methods}

In this section, we describe the methodology for how we acquired the structured Grokipedia dataset and compared it to the English Wikipedia corpus.

\subsection{Building the corpus}
We scraped Grokipedia from 28 to 30 October 2025, using an existing list of article URLs that was published on 27 October 2025 \cite{StefanitGrokipediaurlsDatasets}. The structure of Grokipedia articles is static and somewhat simple, and seems to be derived from Markdown rendering. As such, we were able to build a structured parser for Grokipedia articles (including individual citations). We scraped Grokipedia using parallel processing on Google Cloud, routing requests through a proxy. In total, we were to successfully scrape 883,858 articles from the service, representing 99.8\% of all published Grokipedia articles.

Our corpus of English Wikipedia articles was derived from the daily Wikimedia Enterprise English Wikipedia dump for 28 October 2025, which contains the entirety of English Wikipedia for that day. Once we had a full list of Grokipedia articles, we then joined this dump on title to filter out all articles which did not have an exact title match with a corresponding article on Grokipedia.

\subsection{Embeddings}
Unlike previous work, which compared the official Russian language Wikipedia project with a state-sanctioned alternative called Ruwiki \cite{trokhymovychCharacterizingKnowledgeManipulation2025}, Grokipedia does not use MediaWiki software to run their service. This means we cannot directly track individual edits and editors over time. In order to get a sense of overall article similarity (and which articles have changed significantly), we utilized embedding cosine similarity.

More precisely, for both the Grokipedia and Wikipedia corpora, we extracted the plaintext content of each article in 250-token chunks, with a 100-token overlap between chunks. We then embedded each of these chunks using Google's EmbeddingGemma \cite{veraEmbeddingGemmaPowerfulLightweight2025}, a state-of-the-art 300M parameter embedding model. Once we had embeddings, we calculated the within-article pairwise cosine similarity for each chunk. This allows us to meaningfully discuss metrics like content similarity (filtered by various factors), average article similarity (aggregated across chunks), and more.

\subsection{External data sources}
When elaborating the differences between two corpora, it is important to ask: ``Compared to what?'' We join our datasets on a variety of external sources to evaluate extra-textual features:

\shortsection{Article popularity} The Wikimedia Foundation (WMF) publishes statistics about total human pageviews on an hourly, daily, and monthly basis \cite{PageviewsAnalysisMetaWiki}. In order to get a sense of \textit{which} articles Grokipedia decided to launch with, we use WMF's October 2025 monthly pageview aggregations. To get a sense of relative geographic popularity, we utilize a month of WMF's geo-pageview dataset \cite{adeleyePublishingWikipediaUsage2025}, which reports the countries in which individual pageviews occurred in a privatized manner.

\shortsection{Source reliability} To measure changes in the composition of citation reliability, we use two datasets. The first is the English Wikipedia Perennial Sources list \cite{WikipediaReliableSources2025}, a community-maintained, ``non-exhaustive list of sources whose reliability and use on Wikipedia are frequently discussed'' in English Wikipedia. This dataset gives us a sense of how Grokipedia's citations differ from English Wikipedia's editorial norms. We also utilize a list of crowd-sourced, real-valued domain reliability ratings from 0 to 1, published by Lin et al.~\cite{linHighLevelCorrespondence2023}. This dataset gives us a sense of broad public perceptions of citation reliability for both corpora.

\shortsection{Congresspeople and Members of Parliament subset} We derive the Congresspeople and MPs subset using Wikidata. Specifically, we query Wikidata for people who are current US Senators, members of the US House of Representatives, or members of the 59th Parliament of the United Kingdom, along with their parliamentary party.

\shortsection{Controversial article subset} We derive the controversial article subset using English Wikipedia's existing ``Wikipedia controversial topics'' article category \cite{CategoryWikipediaControversial2023}. Using the public English Wikipedia API, we scrape 3,207 controversial articles. Then, we filter out any non-article pages (files, user pages, category pages, etc.) and remove extraneous prefixes and suffixes (e.g., ``Talk:'' or ``/Archive X''), yielding 3,152 controversial article titles. Finally, we join those titles on Grokipedia titles, yielding 2,056 articles for analysis.

\shortsection{Topics and article quality} We further break down shifts from Wikipedia to Grokipedia by taking a random subsample of 30,000 articles and querying WMF's existing article topic \cite{johnson2021classification} and article quality \cite{johnson2022quality} models. This gives us a set of predicted article topics (at varying levels of specificity) and the predicted article quality class.

\subsection{Other methodological choices}
We measure whether a given Grokipedia article is CC-licensed by doing a simple string search for presence of the footer text at the bottom of each CC-licensed article: ``The content is adapted from Wikipedia, licensed under Creative Commons Attribution-ShareAlike 4.0 License.''

%% file: 03_findings.tex
\section{Findings}

In this section, we summarize findings from our initial work comparing Grokipedia and Wikipedia.

\subsection{Grokipedia corpus characteristics}

At the time of our scrape, Grokipedia stated that there were 885,279 articles on the site, roughly 12\% of the number of articles available on English-language Wikipedia \cite{WikipediaSizeWikipedia2025}. The selection of this subset appears to have been informed primarily by readership. Using WMF's public monthly aggregations of pageview data, we can see that Grokipedia generally contains popular articles: equivalent entries on English Wikipedia accounted for over 4.9 billion non-bot pageviews in October 2025, or roughly 69\% of site pageviews that month. The geolocated pageview data from the month prior to Grokipedia's release shows that the geographic distribution of traffic to articles that exist on Grokipedia skews heavily toward the English-speaking world, with traffic from the US, UK, India, Canada, and Australia accounting for 95\% all pageviews (see Appendix Figure~\ref{fig:pageview_map}).

496,058 of the Grokipedia entries that we were able to scrape displayed a Creative Commons Attribution-ShareAlike license (56\% of the total) while 385,139 do not. This distinction is significant when we review the similarity of the two services: entries with the CC license are far more similar to Wikipedia articles for the same topic than the ones where we did not crawl an equivalent license, suggesting the two groups may have been created or edited with different prompts from Grokipedia's creators. The user interface of the site also reflects this potential methodological difference: CC-licensed articles on Grokipedia contain a public log of edits that Grok made to the source Wikipedia article, and non-CC-licensed articles do not. We were unable to scrape this information on our first attempt, and capturing and analyzing this data remains an area for future work.

\begin{figure}
    \centering
    \includegraphics[width=0.8\linewidth]{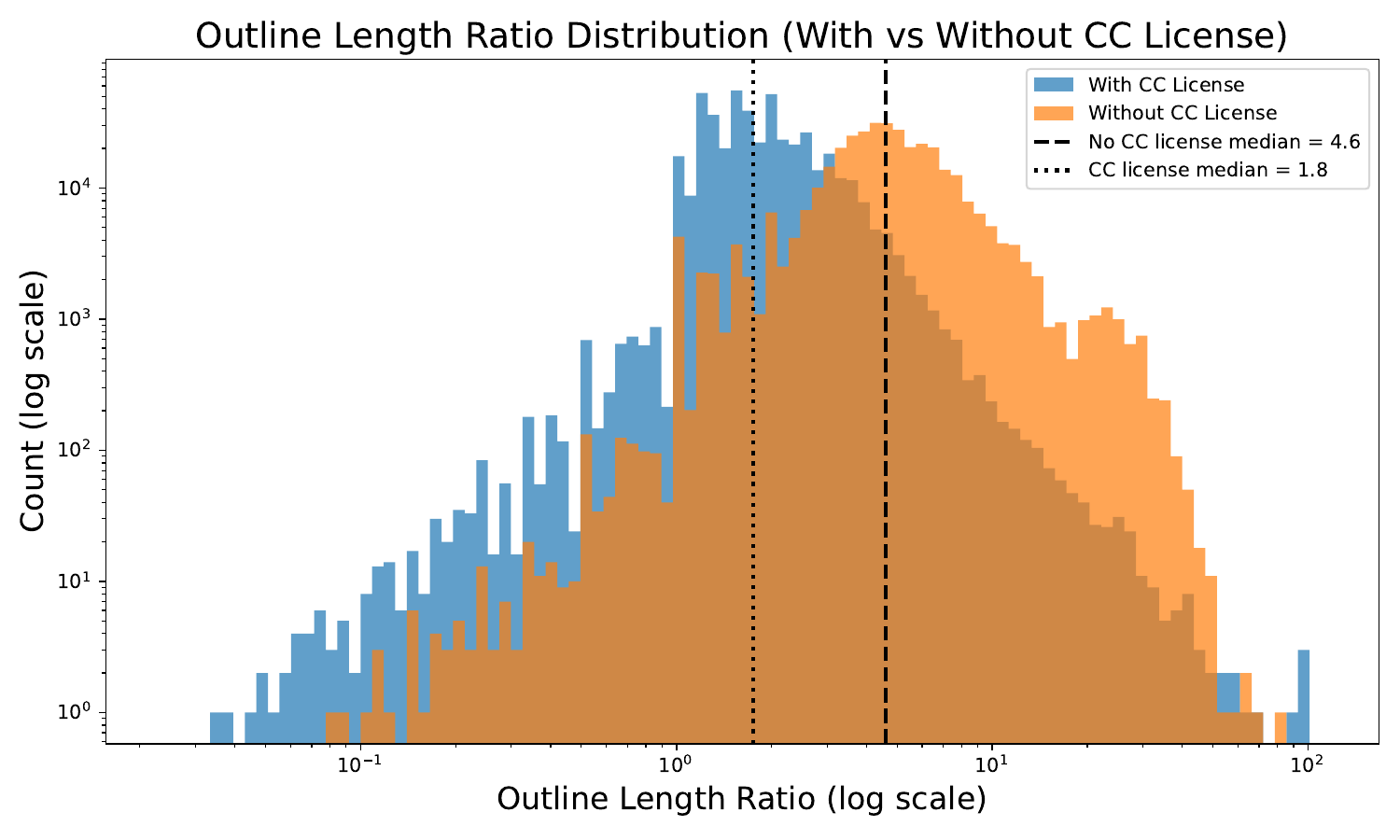}
    \caption{The distribution of article outline length ratios, split by Grokipedia article license status.}
    \label{fig:article-length-ratios}
\end{figure}

Grokipedia articles are significantly longer than their corresponding Wikipedia counterparts. Approximately 96\% of Grokipedia contain as many or more text chunks than their Wikipedia counterparts. Similarly, if we parse out article structure and measure article length in terms of its outline structure, we can see that the median non-CC-licensed Grokipedia article is approximately 4.6 times longer than its Wikipedia counterpart, and some Grokipedia articles are dozens of times longer (see Figure~\ref{fig:article-length-ratios} ).

\subsection{Similarity}
\label{subsec:similarity}

\begin{figure}
    \centering
    \begin{minipage}{0.49\textwidth}
        \centering
        \includegraphics[width=0.98 \linewidth]{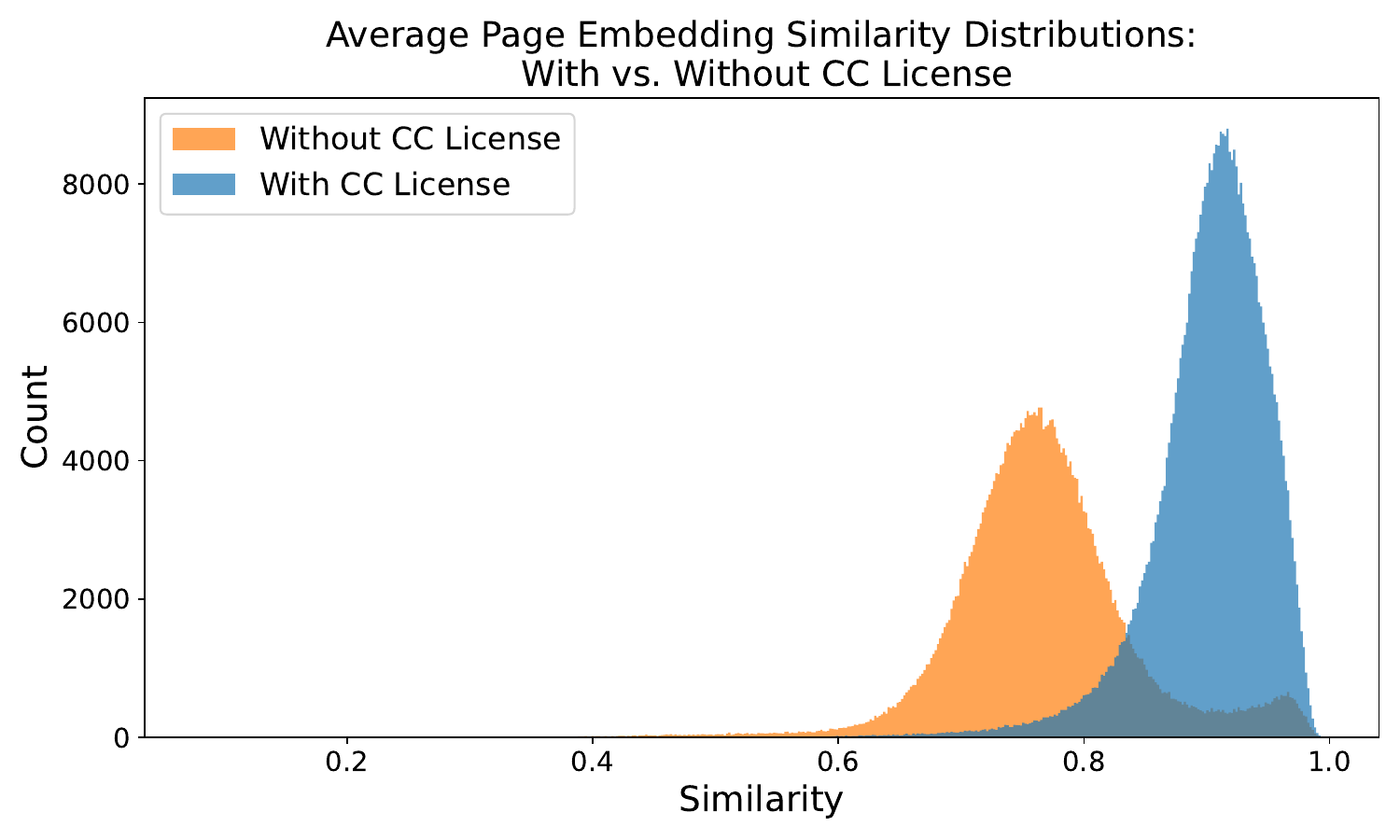}
        \caption{Per-article average similarity embedding distributions, split by whether the Grokipedia article is CC-licensed or not.}
        \label{fig:overall_embedding_sim}
    \end{minipage}
    \hfill
    \begin{minipage}{0.49\textwidth}
        \centering
        \includegraphics[width=0.98\linewidth]{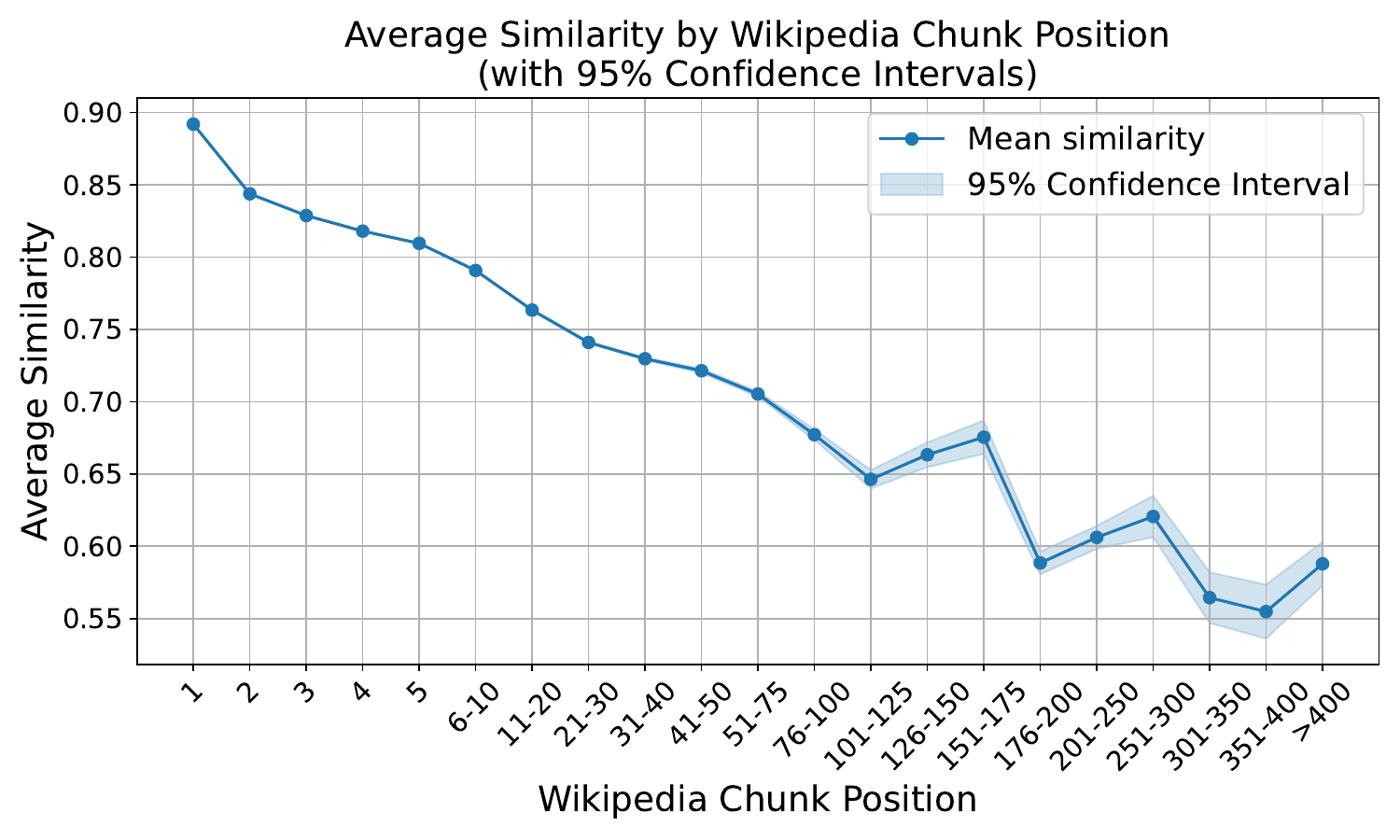}
        \caption{Average chunk similarity by position in an article. Position 1 is the beginning of an article, position 2 is the second chunk, etc.}
        \label{fig:chunk_position_sim}
    \end{minipage}
\end{figure}

As Figure~\ref{fig:overall_embedding_sim} shows, non-CC-licensed entries on Grokipedia follow a bimodal distribution with a mean chunk similarity to their Wikipedia equivalents of 0.77. The similarity for entries with the license is more heavily distributed towards the far end of the spectrum, with a much higher mean chunk similarity of 0.90. We do not know precisely why the non-CC-licensed entry distribution shows bimodality, but speculate that the higher peak corresponds to shorter non-CC-licensed articles. This may provide an explanation because, across both corpora, per-chunk similarity is roughly dependent on chunk position in an article (see Figure~\ref{fig:chunk_position_sim}). Average chunk similarity is highest in the introduction and decreases roughly linearly as the article proceeds. This finding is not attributable to changes in article structure, as we calculated pairwise similarity for each chunk on Wikipedia with each chunk on its equivalent Grokipedia entry and compared the two most similar ones.

Grokipedia articles with very high average chunk similarity to their corresponding Wikipedia article include verbatim transcriptions. These articles appear in both the CC-licensed and non-CC-licensed subsets of the data; that is, identical articles (or chunks) do not necessarily carry an attribution to Wikipedia or a CC license. For instance, Table~\ref{tab:high_sim_comp} shows two excerpts from Grokipedia entries that have exact matches on equivalent Wikipedia articles. The entry for the Mejia Thermal Power Station is not CC-licensed, whereas the one for Sono Sachiko, a 19th century member of the Japanese imperial family, attributes Wikipedia. Note that, in the non-CC-licensed Mejia Thermal Power Station page, the first sentences on both Wikipedia and Grokipedia include the same \textit{typo}: ``Commissioned on [sic] 1996''. More excerpt pairs from various parts of the similarity distribution are available in Appendix~\ref{app-subsec:comp}. In some cases, dissimilarity is to Grokipedia readers benefit. The entry for Canadian actress Ellen David on Wikipedia includes a ``Career'' section that is a very sentence listing the movies she has starred in; the AI-written entry is a more readable summary, with her roles summarized in tabular format below (see \ref{app-subsec:comp}).

\begin{table}[ht]
    \centering
    \renewcommand{\arraystretch}{1.2}
    \setlength{\tabcolsep}{6pt}
    \begin{tabular}{|p{3.2cm}|p{6cm}|p{6cm}|}      \hline
        \textbf{Entry} & \textbf{Wikipedia} & \textbf{Grokipedia} \\
        \hline
        \textbf{Sono Sachiko (CC licensed)} &
        Sachiko's father was Count Sono Motosachi; she was known as Kogiku Tenji. She gave birth to two sons and six daughters, several of whom died prematurely... &
        Sachiko's father was Count Sono Motosachi; she was known as Kogiku Tenji. She gave birth to two sons and six daughters, several of whom died prematurely... \\
        \hline
        \textbf{Mejia Thermal Power Station (Not CC licensed)} &
        Mejia Thermal Power Station is located at Durlabhpur, Bankura, 35 km from Durgapur city in West Bengal. The power plant is one of the coal based power plants of DVC. Commissioned on 1996... &
        Mejia Thermal Power Station is located at Durlabhpur, Bankura, 35 km from Durgapur city in West Bengal. The power plant is one of the coal based power plants of DVC. Commissioned on 1996... \\
        \hline
    \end{tabular}
    \caption{Comparison of Wikipedia and Grokipedia entries.}
    \label{tab:high_sim_comp}
\end{table}

\subsection{Top 100 sources}
\label{subsec:t100}
The top 100 domains cited by Wikipedia and Grokipedia account for respectively 30.4\% and 25.2\% of their overall sources. 57 of these domains are the same, even if their precise placement among the top 100 varies. Frequently-cited domains across both services include media outlets like nytimes.com, theguardian.com, and bbc.com; online portals like archive.org, and books.google.com; and specialized databases like imdb.org and baseball-reference.com.

\begin{figure}
    \centering
    \includegraphics[width=.68\linewidth]{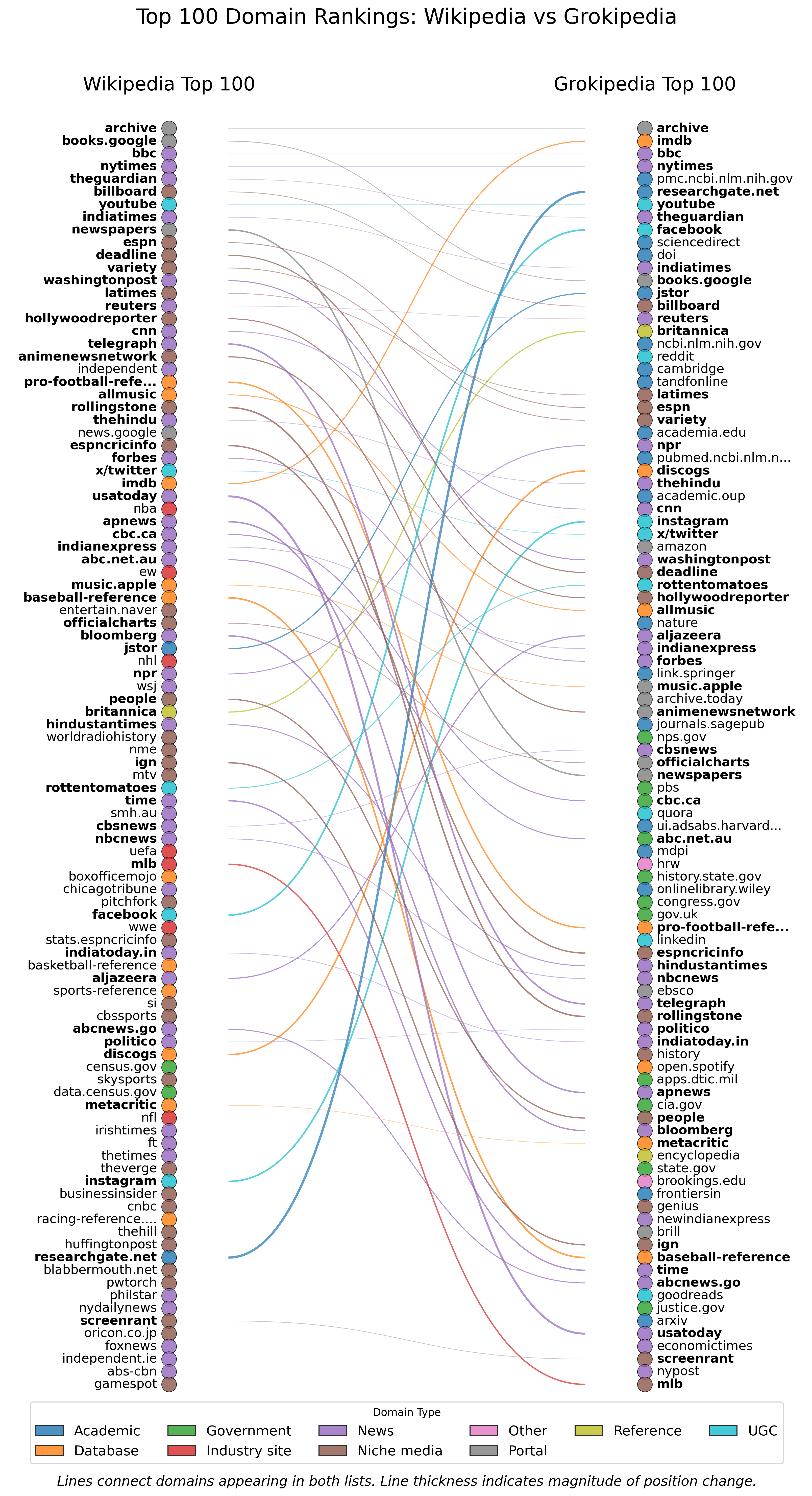}
    \caption{The top 100 most-cited domains on Wikipedia and Grokipedia. Domains are bold if they are on both lists, and lines connecting the domains show their position change. Color indicates domain type.}
    \label{fig:t100-pos-comp}
\end{figure}

However, there are distinctions in the types of sources most often cited by the two encyclopedias. Some of these appear to be a consequence of Wikipedia's policy preference for secondary sources via the ``No Original Research'' policy \cite{WikipediaNoOriginal2025}. Wikipedia's top 100 includes 40 mainstream media outlets compared with Grokipedia's 29. Only two academic sources---jstor.org and researchgate.net---make it into the Wikipedia's most-cited sources, whereas the AI encyclopedia's top 100 features 19. Grokipedia also relies more heavily on social networks and other UGC platforms. Websites like Facebook, Reddit, Quora, YouTube, Amazon, and Musk's own social network X account for nine of Grokipedia's top 100 sources (and 2.8\% of all citations), compared to 5 sources and 1.5\% of citations for Wikipedia. Figure~\ref{fig:t100-pos-comp} shows the top 100 most-cited sources in both corpora, along with domain type and ranking changes. Appendix Figure~\ref{fig:t100-treemap} shows an area map of top 100 source proportion in both corpora.

Notably, although 56\% of Grokipedia articles contain licensing text indicating that they are adapted from Wikipedia and much of the corpus displays meaningful textual similarity to Wikipedia, Wikimedia Foundation domains (wikipedia.org, wiktionary.org, wikisource.org, wikidata.org, etc.) appear only 445 times, accounting for approximately 1 in 100,000 citations on Grokipedia.

\subsection{Source quality and variance}

Sourcing is a highly contested and regimented aspect of Wikipedia's operations. In this section, we dive into different metrics of source quality, and pull out some curious anecdotes about what Grokipedia cites. Broadly speaking, we find that Grokipedia cites more sources of all kinds (perhaps due to its wordier entries). This includes sources on the far low end of the source reliability spectrum as measured both by in-group English Wikipedia editorial norms and more broad external judgments of domain quality.

\subsubsection{Perennial sources}
The English Wikipedia community maintains the ``Perennial Sources'' list \cite{WikipediaReliableSources2025}. This is a list of sources whose reliability has been frequently discussed on the platform, grouped as ``generally reliable,'' ``generally unreliable,'' ``blacklisted,'' ``no consensus,'' and ``deprecated.'' This list is not comprehensive---it includes only 880 domains---and it is not strictly enforced. Inclusion in the ``generally unreliable'' (or even ``blacklisted'') source category doesn't mean a domain is in practice never cited on Wikipedia, but it is discouraged. Nevertheless, the Perennial Sources list has drawn extensive media attention, especially in the United States \cite{cohenWhyWikipediaDecided, tuckercarlsonWikipediaCoCreatorReveals2025}. In this paper, we utilize this list as a means of (roughly) approximating within-community citation norms.

\begin{figure}
    \centering
    \includegraphics[width=\linewidth]{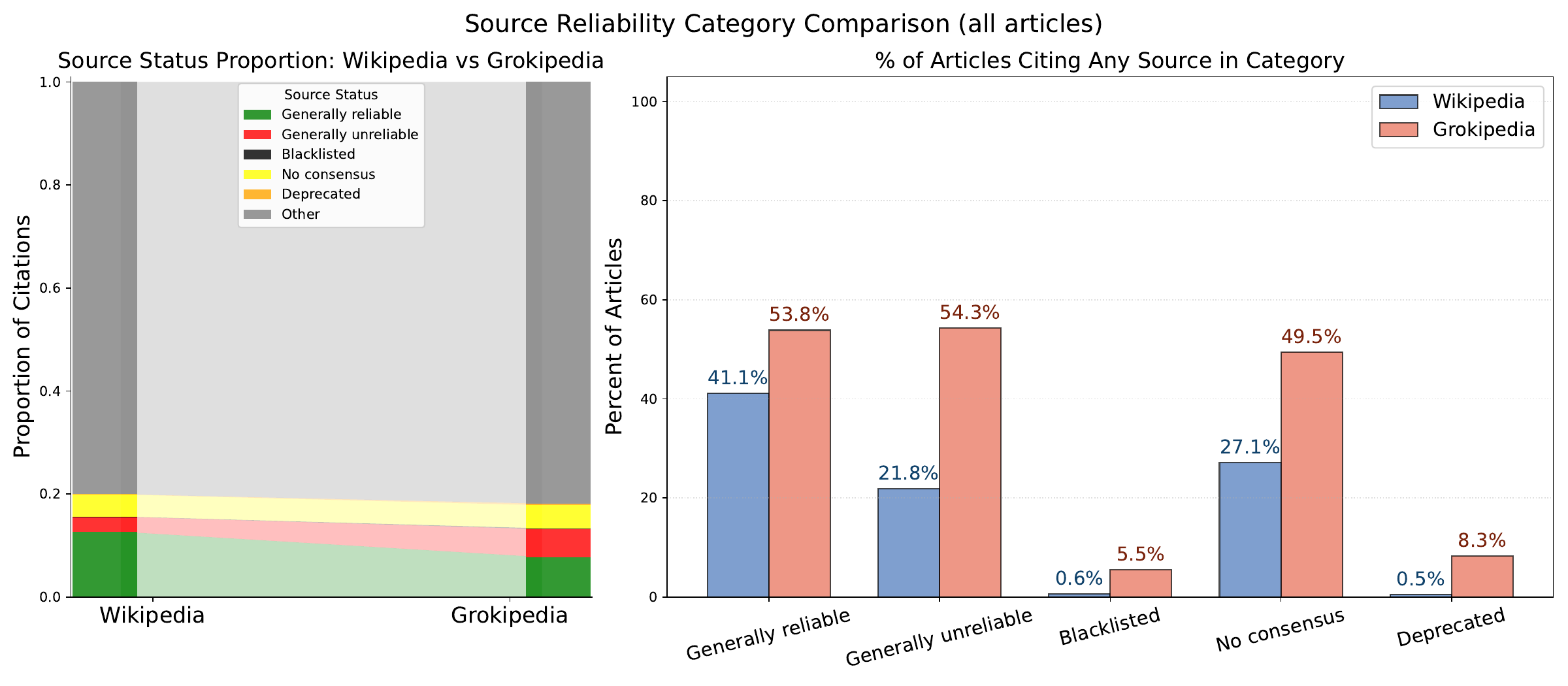}
    \caption{(Left) The relative proportion of Perennial Source list categories in Wikipedia and Grokipedia citations. (Right) The percentage of articles that contain at least one of each category of citation in the two corpora.}
    \label{fig:overall-grok-wp-cite-comp}
\end{figure}

Perhaps as expected, given the Perennial Sources list's prominence in the English Wikipedia community, ``generally reliable'' sources make up a far larger proportion of Wikipedia citations (12.7\%) than ``generally unreliable'' (2.9\%) or ``blacklisted'' (0.04\%) sources (see Figure~\ref{fig:overall-grok-wp-cite-comp}). ``Generally reliable'' sources are cited in roughly 2 of 5 (41.1\%) of Wikipedia articles, as opposed to roughly 1 in 5 (21.8\%) articles citing ``generally unreliable'' sources and 1 in 167 (0.6\%) articles citing ``blacklisted'' sources. Grokipedia's citation practices appear to be less in line with Wikipedia editorial norms. ``Generally reliable'' sources make up 7.7\% of citations on Grokipedia (a relative decrease of 39\%), ``generally unreliable'' sources are 5.4\% of citations (a relative increase of 86\%), and ``blacklisted'' sources make up 0.1\% of citations (a relative increase of 275\%). At the article level, the increase is even more drastically visible: 5.5\% of Grokipedia articles contain at least one citation to ``blacklisted'' sources---a ninefold increase in prevalence compared to Wikipedia.

This comparison, however, is across the entire Grokipedia corpus, without distinction along the lines of article licensing that we saw was so important in Section~\ref{subsec:similarity}. When we look at just the subsets of the Grokipedia corpus that do not have a CC-license (and their corresponding Wikipedia articles), the increase in articles that cite a ``blacklisted'' source is even more dramatic: 0.9\% of Wikipedia articles and 11.7\% of Grokipedia articles, a 13x increase (see Appendix Figure~\ref{fig:overall-grok-wp-cite-comp-no-cc}).

\subsubsection{Domain reliability scores}

\begin{figure}
    \centering
    \includegraphics[width=1\linewidth]{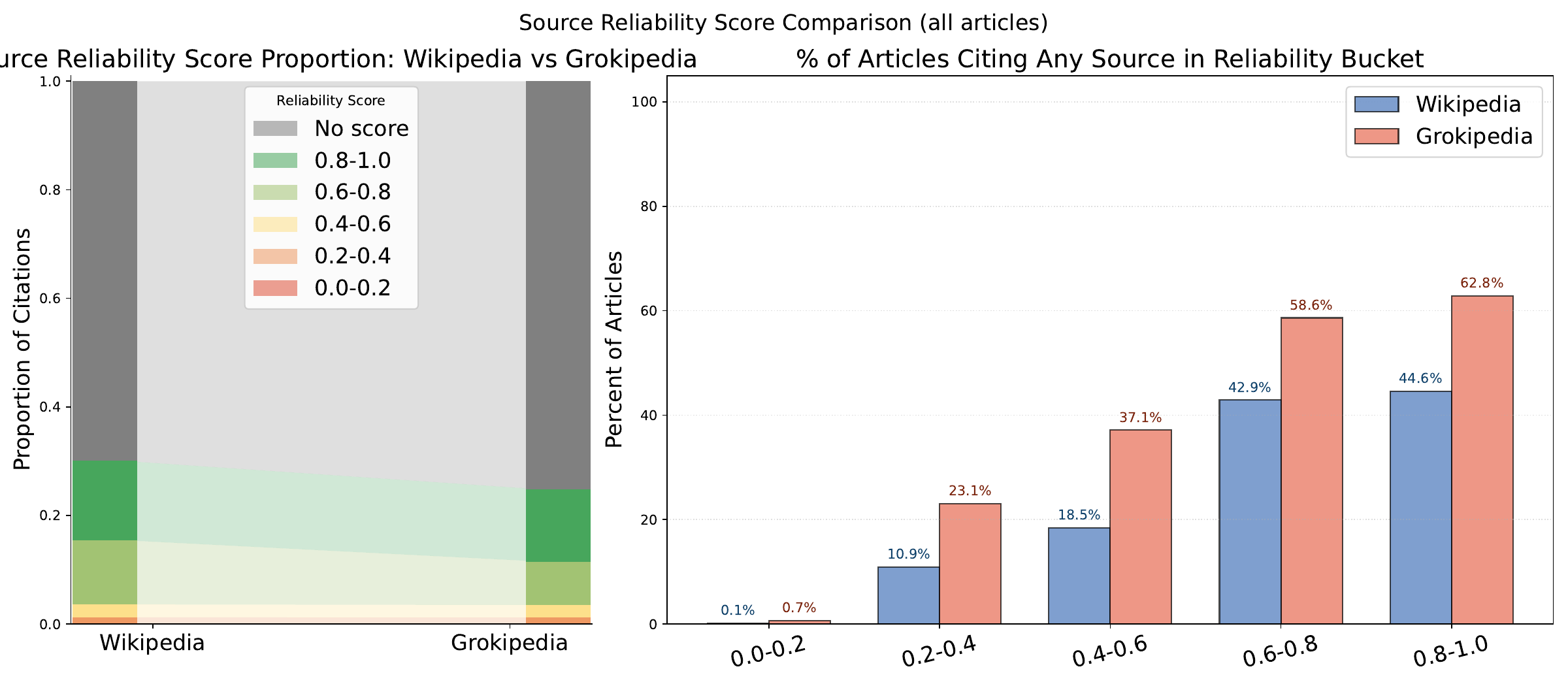}
    \caption{(Left) The relative proportion of Lin et al. score reliability in Wikipedia and Grokipedia citations. (Right) The percentage of articles that contain at least one domain in each score bucket in the two corpora.}
    \label{fig:overall-grok-wp-cite-lin}
\end{figure}

To get a sense of source quality that isn't exclusively tied to English Wikipedia's editorial norms, we use Lin et al.'s domain reliability scores \cite{linHighLevelCorrespondence2023}. These are real-valued scores from a low of 0.0 to a high of 1.0, generated via combining various expert news rating sources. Our findings on source quality remain roughly in line with those that relied on English Wikipedia's Perennial Source list (see Figure~\ref{fig:overall-grok-wp-cite-lin}): English Wikipedia is more likely to cite domains on the 0.6 and above higher end of Lin et al.'s quality score (27.4\% of all citations) than Grokipedia (21.3\% of all citations). Low quality domains---which we define as having quality scores between 0.0 and 0.2---make up three times the share of total citations on Grokipedia than Wikipedia. Even though this share is relatively small (0.03\% of the total), it means Grokipedia includes 12,522 citations to domains deemed of very low credibility. Websites in this category include the white nationalist forum Stormfront, the antivaccine conspiracy website Natural News, and the fringe website InfoWars. None of these domains are cited at all on Wikipedia; they have 180 citations on Musk's service. Although Grokipedia quotes more sources per article in all quality buckets, it disproportionately adds low quality sources to articles.

Unsurprisingly, source quality is especially affected in the subset of Grokipedia articles that do not have the CC license and which are therefore more dissimilar from their Wikipedia equivalents (see Appendix Figure~\ref{fig:overall-grok-wp-cite-lin-no-cc}). This suggests that, given free rein to cite sources, Grokipedia is less discerning with source quality than Wikipedia.

A limitation with both source quality scores is that they don't rate the majority of citations used on either service. What we can say at this stage is that Grokipedia is both more capacious in its citations---almost doubling Wikipedia's total---and more ecumenical in its approach, including many more sources across all quality buckets.

\subsubsection{Citation curiosities}

We found a variety of curiosities as we dug into Grokipedia's citation practices, each of which might warrant further study.

\shortsection{Rarely cited domains on Wikipedia} We reviewed the 1,071 domains for which we have a Lin et al. quality score that were cited on Grokipedia but received at most 1 citations on Wikipedia. The mean quality score of these domains is 0.48. Besides the aforementioned InfoWars and Stormfront, this list of domains includes American and Indian right-wing media outlets (swarajyamag.com, 6,369 citations; republicworld.com, 1,330; breitbart.com, 566); Chinese and Iranian state media (globaltimes.cn, 2,903; presstv.ir, 778); anti-immigration, anti-Semitic or anti-Muslim websites (unz.com, 227; vdare.com, 107; frontpagemag.com, 91; jihadwatch.org, 73) and websites that have been variously accused of promoting pseudoscience and conspiracy theories (ancient-origins.net, 7665; lifesitenews.com, 100; thegatewaypundit.com, 90; globalresearch.ca, 51; voltairenet.org, 45).

Grokipedia cites these sources without qualifying their reliability. The Grokipedia article ``Clinton body count,'' for example, cites InfoWars reporting twice (as citations ([128] and [129]):
\begin{quote}
    InfoWars, for instance, has published articles compiling deaths of individuals associated with the Clintons, such as former aide Mark Middleton, whose 2022 suicide by gunshot—despite being found with an extension cord around his neck and tied to a tree—prompted coverage questioning official rulings [128]. These compilations, often updated with new cases, frame patterns of timing and circumstances as non-coincidental, drawing on public records and witness accounts to argue for scrutiny beyond mainstream dismissals [129].
\end{quote}

\shortsection{X/Twitter usernames} X/Twitter received approximately 40,000 more citations on Grokipedia than on Wikipedia. We thought it may be interesting to parse \textit{which accounts} Grokipedia cites from X/Twitter. Grokipedia cites \texttt{@grok} 232 times (as opposed to 0 on Wikipedia) and \texttt{@elonmusk} 186 times (as opposed to 49 on Wikipedia). We also found one article (``Sarah Anne Williams'') where the X/Twitter account of the \textit{subject of the article} (\texttt{@SarahAnneWillia}) was cited 99 times.

\shortsection{Grok conversations} As we were parsing X/Twitter user data, we found 1,050 citations to links containing the string \texttt{\{twitter, x\}.com/i/grok/share}. These links are to publicly shared conversations between X users and the Grok chatbot, and are interesting to dig into. For instance, the first section the Grokipedia article for ``Guy Verhofstadt'' contains this sentence:
\begin{quote}
    His tenure as prime minister occurred amid challenges including the 1999 dioxin contamination scandal's aftermath, which affected public trust in Belgian institutions, though his government focused on recovery and stability measures [7].
\end{quote}

Citation 7 is a link to a Grok chatbot conversation on X, where the user starts by asking Grok: ``Can you dig up some dirt on Guy Verhofstadt please?'' Grok provides an extensive answer, including the claim that makes it into the beginning of the article. Grok ends its answer with ``The dirt’s there—some solid, some speculative.''  

Between hundreds of citations to the \texttt{@grok} X/Twitter account and approximately a thousand links to Grok conversations, Grokipedia may be engaging in a new kind of ``LLM auto-citogenesis''\footnote{\url{https://xkcd.com/978/}}: the user-directed creation of LLM text on third party sites, which then are utilized \textit{by the same LLM} (or one in the same model family) to write an encyclopedic resource. This analysis of citation composition on the two sites is just scraping the surface. We believe there is much future work to complete in this area.

\subsection{Political figures}

\begin{figure}
    \centering
    \includegraphics[width=1\linewidth]{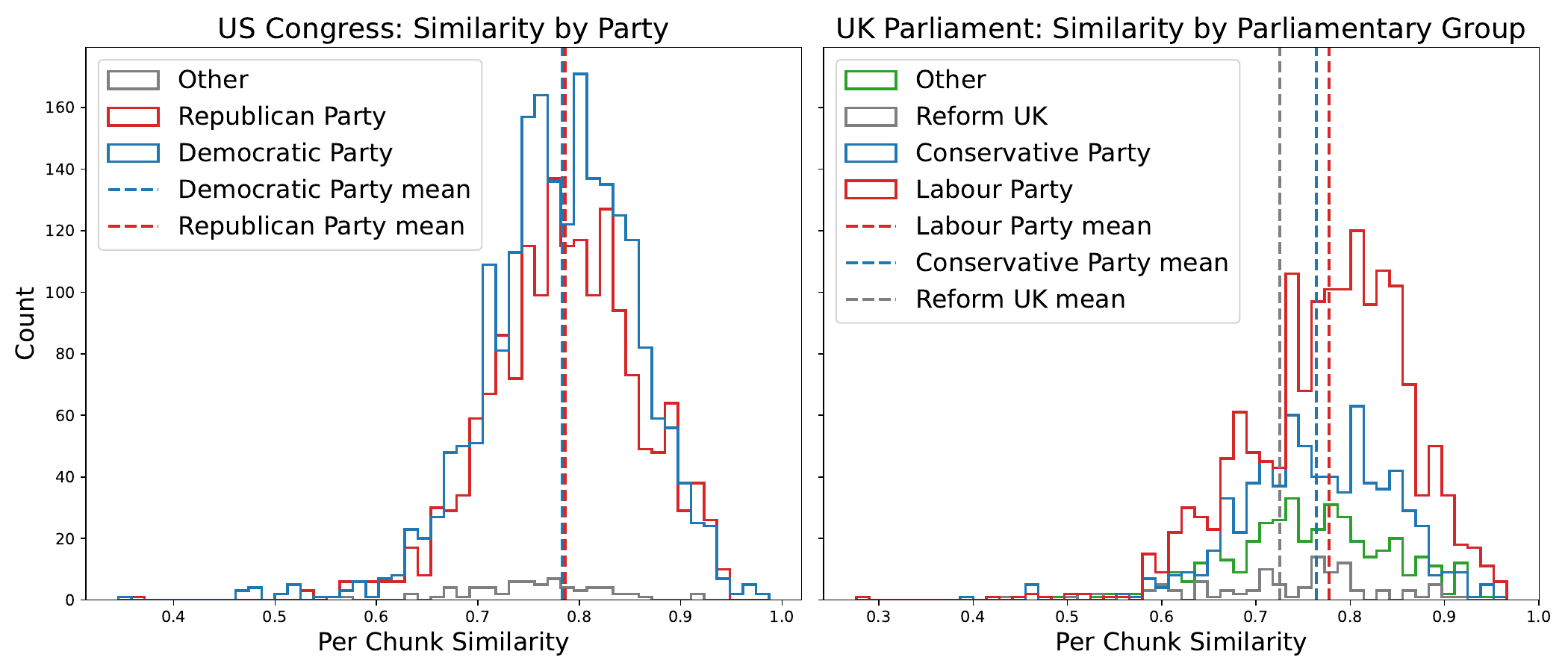}
    \caption{US and UK elected official article similarity by party.}
    \label{fig:us-uk-party-sim}
\end{figure}

Moving from Grokipedia as a whole to entries for more sensitive verticals, like elected officials, the discrepancy in source quality with Wikipedia holds true. For both members of the US Congress and the UK Parliament, Grokipedia's share of higher quality citations tends to be lower than that of Wikipedia's and that of lower quality sources tends to be higher. For both UK and US elected officials, average chunk similarity is below 0.8, with some inter-party variation in the United Kingdom (Figure~\ref{fig:us-uk-party-sim}). 

Reviewing two highly dissimilar entries in this subcategory---the articles for UK Prime Minister Keir Starmer and Seattle Congresswoman Pramila Jayapal---we can see that Grokipedia is more prominent and vocal in detailing alleged controversies surrounding the two politicians.

Jayapal's Grokipedia entry mentions in the introduction that she ``has faced notable controversies, including backlash for describing Israel as a 'racist state' during a 2023 event.'' Starmer's article states that his tenure as Director of Public Prosecutions ``drew criticism for low prosecution rates in child sexual exploitation cases during the emergence of grooming gang scandals and decisions not to pursue charges in high-profile matters such as the initial handling of complaints against Jimmy Savile.'' (This particular accusation is one that Elon Musk himself has leveled against Starmer \cite{coureaWhyElonMusk2025}.)

Anecdotally, Grok edits seem to increase the prominence and frequency of references to the subject's bias or controversies. Additional work is required to determine whether this is true across the entire corpus.

\subsection{Controversial topics}

Articles in Wikipedia's list of controversial topics \cite{CategoryWikipediaControversial2023} are worth reviewing as a dedicated subset because they are likely to disproportionately reflect Elon Musk's grievances with the crowdsourced encyclopedia. These are articles that are more vulnerable to ``edit wars,'' and, as a result, likelier to turn on edit protections (limiting the population of possible editors to only those with more experience) or have strict guidelines for how edits can be made.

\begin{figure}
    \centering
    \includegraphics[width=.8\linewidth]{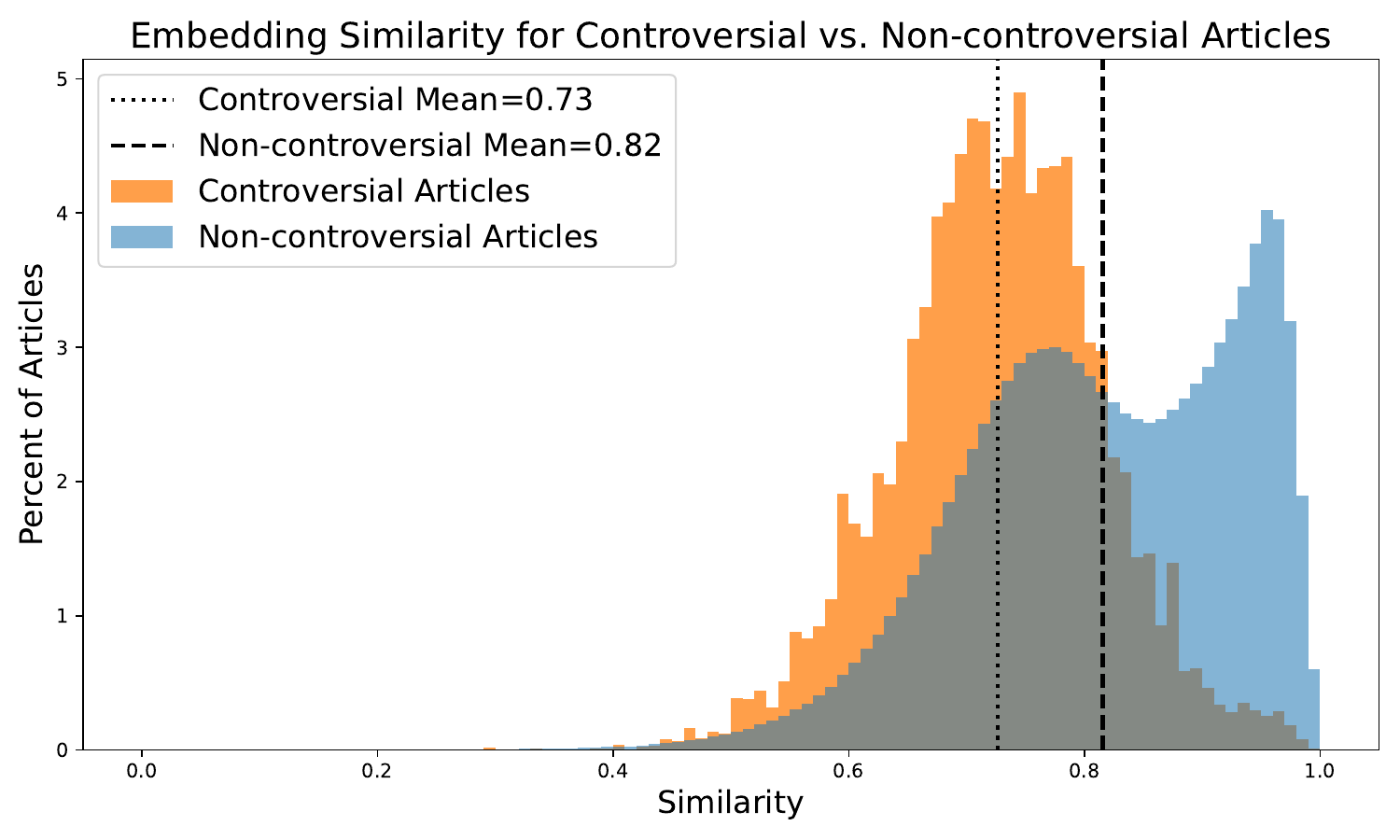}
    \caption{Similarity distributions of controversial article chunks and non-controversial article chunks. Note that the Y-axis of this graph is expressed in terms of percentage of articles (as opposed to absolute counts).}
    \label{fig:controversial-similarity}
\end{figure}

Indeed, we find that Grokipedia entries that map to a article in the controversial category are on average far more dissimilar than other articles from their Wikipedia equivalents (mean similarity = 0.73) compared to those that do not map to an article in the controversial category (mean similarity = 0.82) (Figure~\ref{fig:controversial-similarity}). Fully 93.1\% of these articles (1,914 out of 2,056 articles) lack the CC license that we hypothesize appears on articles that Grok was instructed to recreate more faithfully from Wikipedia, compared to the global proportion of 56\%.

A randomly selected sample of highly dissimilar articles in the controversial bucket include the following entries: ``Criticism of the United States government,'' ``Racism in the United States,'' ``Media bias in the United States,'' ``Masculism,'' and ``Comparative Studies of the Roman and Han empires.'' The edits themselves suggest an ideological bent in the writing instructions, or at minimum an attempt to review the original text from a different perspective. The Wikipedia entry for ``Masculism'' amply cites scholars that view the movement as anti-feminist or misogynist; the Grokipedia entry states that ``its core claims draw on verifiable statistics rather than ideology alone.'' The Wikipedia entry for ``Media bias in the United States" starts with a sweeping historical review of the American information ecosystem that is perhaps overly critical of the arrival of the Internet. The Grokipedia entry is much more concerned with what it calls ``empirically observed left-liberal slant in the news coverage of major mainstream outlets'' (see Table~\ref{tab:media-bias}). Nonetheless, the sample size here is way too small for this interpretation to be anything other than evocative and an invitation for further study.

\begin{table}
    \centering
    \renewcommand{\arraystretch}{1.2}
    \setlength{\tabcolsep}{6pt}
    \begin{tabular}{|p{7cm}|p{7cm}|}
        \hline
        \textbf{Wikipedia} & \textbf{Grokipedia} \\
        \hline
        The history of media bias in the United States has evolved from overtly partisan newspapers in the 18th and 19th centuries to professional journalism with ethical standards in the 20th century, and into the 21st century, where the Internet enabled anyone to become a journalist. As the public stopped paying for news, socially responsible journalism became difficult to sustain, and the floodgates opened to individuals lacking formal training who could publish stories with little effort or expertise. &
        Media bias in the United States refers to the empirically observed left-liberal slant in the news coverage of major mainstream outlets. This bias arises from the ideological imbalance within the journalism profession and results in disproportionate emphasis on progressive narratives, selective fact presentation, and unfavorable portrayal of conservative viewpoints. \\
        \hline
    \end{tabular}
    \caption{Comparison of Wikipedia and Grokipedia descriptions of media bias in the United States.}
    \label{tab:media-bias}
\end{table}

\begin{figure}
    \centering
    \includegraphics[width=.8\linewidth]{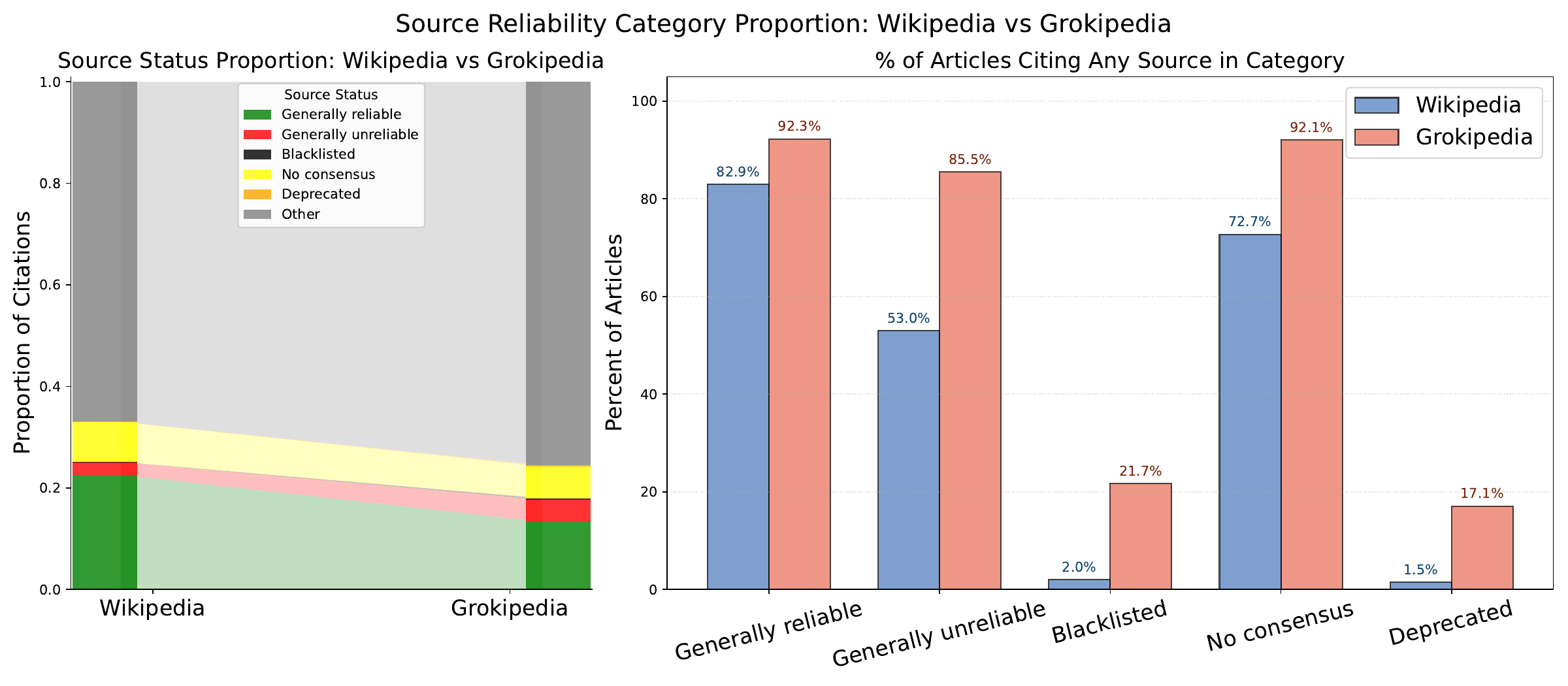}
    \caption{(Left) The relative proportion of Perennial Source list categories in Wikipedia and Grokipedia citations for controversial articles. (Right) The percentage of controversial articles that contain at least one of each category of citation in the two corpora.}
    \label{fig:controversial-cite-comparison}
\end{figure}

\begin{figure}
    \centering
    \includegraphics[width=.8\linewidth]{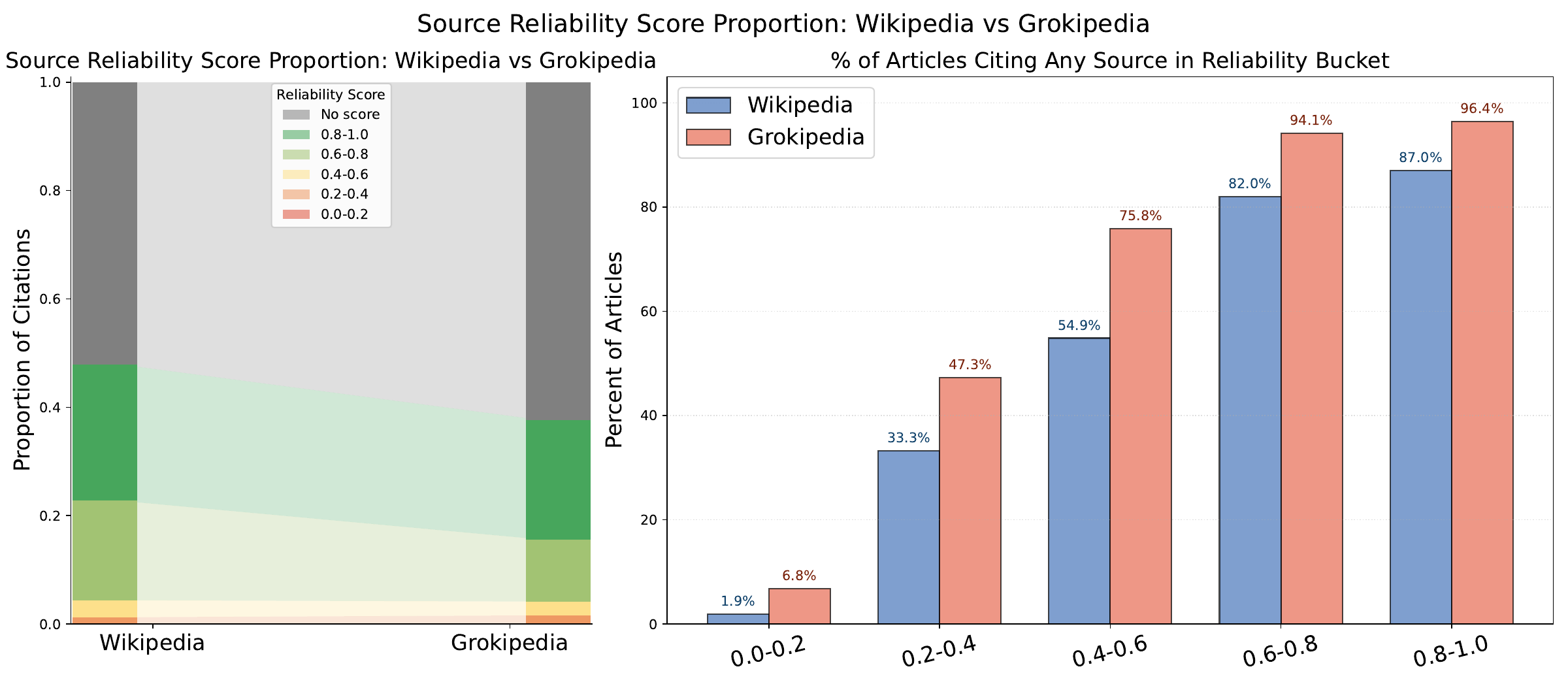}
    \caption{(Left) The relative proportion of Lin et al. score buckets in Wikipedia and Grokipedia citations for controversial articles. (Right) The percentage of articles that contain at least one of each score bucket of citation in the two corpora.}
    \label{fig:controversial-cite-comparison-lin}
\end{figure}

Within these controversial articles, the increase in lower quality sources is particularly visible. By Wikipedia's definition (the Perennial Source list), 85.5\% of Grokipedia's entries on controversial topics contain a citation to a ``generally unreliable'' source and 21.7\% contain a citation to a source that has been ``blacklisted'' (see Figure~\ref{fig:controversial-cite-comparison}). The prevalence of articles with at least one source from the lowest quality domains in Lin et al. also increases from 1.9\% on Wikipedia to 6.4\% on Grokipedia (see Figure~\ref{fig:controversial-cite-comparison-lin}).


\subsection{Article quality and topic}

As mentioned in Section~\ref{sec:methods}, we derived article quality and topic for a randomly selected subset of 30,000 English Wikipedia articles using publicly accessible WMF models.

\shortsection{Article quality} WMF provides a model that, given an article title, predicts the \textit{quality class} of that article \cite{johnson2022quality}. Wikipedia articles are categorized by their quality: for example, an article with a ``Featured Article'' (FA) designation could be featured on the front page of Wikipedia. Other quality classes include: ``Good Articles'' (GA), ``Starts,'' ``Stubs,'' and quality ratings like A, B and C.

\begin{figure}
    \centering
    \includegraphics[width=\linewidth]{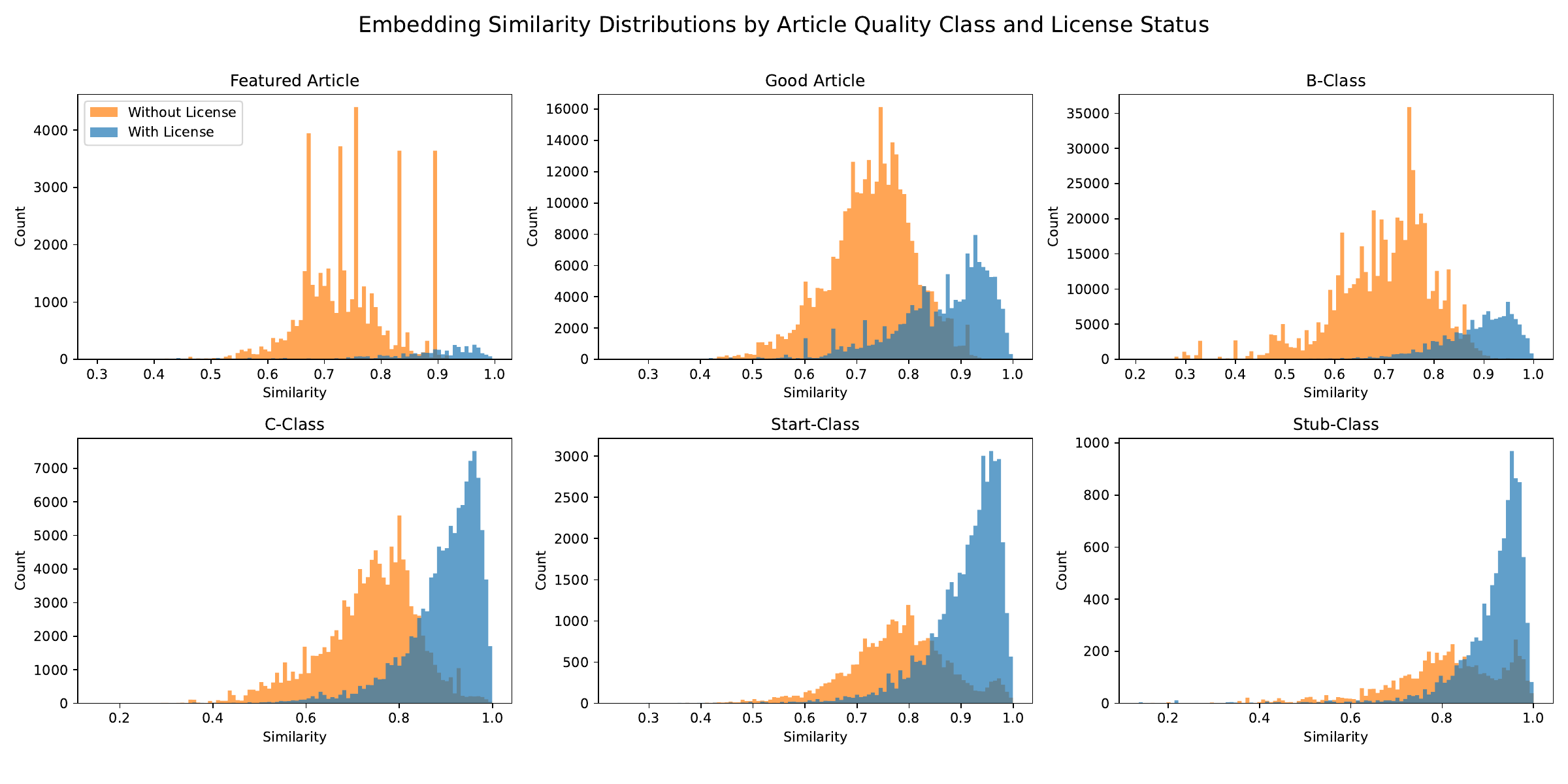}
    \caption{A set of histograms showing chunk similarity between Grokipedia and Wikipedia articles in our random subsample by article class, split by license status.}
    \label{fig:sim-by-quality}
\end{figure}

Figure~\ref{fig:sim-by-quality} depicts a histogram of chunk similarity by article class, split by license status. Chunks for articles in the FA and GA quality classes---the pages that have been deemed highest quality on English Wikipedia---are far more likely to be non-CC-licensed pages (the orange sections of the graphs). As a result, they also show much less similarity to their English Wikipedia counterparts. Start, Stub, and C-class articles, on the other hand, are much likelier to be CC-licensed, directly adapted from English Wikipedia, and display a higher similarity to Wikipedia. 

\shortsection{Article topic} WMF also provides an open source model that, given an article title, predicts likely article topics from an existing topic ontology \cite{johnson2021classification}. This topic ontology is multi-level (e.g., a prediction about article regionality looks like \texttt{Geography.Regions.Asia.Southeast\_Asia}), and encompasses topics generally relevant to encyclopedic content: there are subtopics pertaining to history, politics, biography, science, geography, etc.

Appendix Figure~\ref{fig:topic-comp} shows per-chunk similarity by article subtopic, split by license status. Similarly to the rest of this work, there is a strong dichotomy between CC-licensed Grokipedia articles and non-CC-licensed Grokipedia articles, where non-CC-licensed articles are significantly less similar to Wikipedia than CC-licensed articles. Of particular note in this figure is \textit{which} topics are largely CC-licensed and which are not. Largely licensed subtopics include those in the \texttt{STEM} category, like \texttt{Biology}, \texttt{Computing}, \texttt{Technology} and \texttt{Engineering}, as well as \texttt{Culture} subtopics like \texttt{Internet\_Culture}, \texttt{Media}, \texttt{Sports}, and \texttt{Literature}. More notable, however, are the topics that a largely unlicensed. In the \texttt{Culture} topic, that includes subtopics like \texttt{Biography} and \texttt{Philosphy\_and\_Religion}. In the \texttt{History\_and\_Society} topic, that includes subtopics like \texttt{Business\_and\_Economics}, \texttt{History}, \texttt{Military\_and\_Warfare}, \texttt{Politics\_and\_Government}, and \texttt{Society}. \texttt{Geography.Regions} also shows a strong bias towards non-CC-licensure.

The random subsample of articles that we used to do the above analyses of article quality and topic only encompasses 3.4\% of the Grokipedia corpus. As such, it certainly cannot tell the entire story. However, it may be indicative of broader trends about Grokipedia. We would welcome future work that does a fuller analysis of both of these features. 

%% file: 05_conclusion.tex
\section{Conclusions, limitations, and future work}

In this paper, we provide a first look at how Grokipedia differs from Wikipedia. At its first release, Grokipedia appears to be heavily indebted to Wikipedia, the site it was intended to replace and outclass. Many articles show a high level of similarity across the two corpora, and some articles contain exactly identical copies of text. Other articles, conversely, had notable shifts in content, framing, and tone. We find that Grokipedia is a significantly larger and longer corpus than Wikipedia, including much more text per article---and many more citations. These citations are from domains across the quality spectrum (as measured both within English Wikipedia community norms and more generally), and it is clear that sourcing guardrails have largely been lifted on Grokipedia. This results in the inclusion of questionable sources, and an overall higher prevalence of potentially problematic sources. We found that these trends were particularly notable in subsets of the corpus that pertain to elected officials and controversial topics. We also stumbled across a variety of curiosities when it came to sourcing, including cases of Grokipedia citing Grok chatbot conversations. 

At the beginning of this paper, we set out to answer the questions of whether Grokipedia is a synthetic derivative or an ideological project. Based on our findings, the answer to both of these questions is a (qualified) yes.

\shortsection{Limitations} Because this is an initial comparison, this paper is somewhat limited in its methodology and external data sources. In particular, the domain sourcing lists are highly incomplete and are not objective final judgments on source quality. We try to acknowledge these imperfections explicitly, and treat them as rough proxies for in-group community sourcing norms and broader expert opinion of domain quality on the internet. The other primary measure we use to compare the two corpora is semantic similarity between article chunks. This, too, is an imperfect instrument. In particular, it can miss more subtle changes in linguistic framing that maintain a high level of textual similarity: single words that have been shifted to alter the interpretation of an event, shifts from the active to the passive voice, etc. Identifying and characterizing these more subtle changes is an area of open work.

\shortsection{Future work} We believe this analysis has only barely scratched the surface, and there is much more to analyze in depth. As mentioned above, identifying and characterizing subtle linguistic framing changes is a challenging problem, and one that we hope to tackle next. We also would be intrigued by efforts to scrape and analyze the public log of edits made to underlying Wikipedia content for CC-licensed articles on Grokipedia. Additionally, it may be interesting to look at tone and writing style; English Wikipedia has a ``Neutral Point of View'' policy \cite{WikipediaNeutralPoint2025} and a manual of style to avoid what Wikipedia editors call ``puffery'' or ``peacocking'' within articles \cite{WikipediaManualStyle2025}. Our controversial articles subset is just a starting point; another line of inquiry would be to parse article content and compare articles that have a subsection about article subject ``controversies.'' Finally, there's the question of claim verifiability, which is an open area of research. Given a claim in Wikipedia or Grokipedia and its source citation, can we rate the claim's verifiability? We would be intrigued by research that seeks to answer all of these questions and more.

%% file: 99_appendix.tex
\section{Appendix}

\subsection{Geographic distribution of pageviews}

The map below shows a log-scaled geographic distribution of pageview traffic to English Wikipedia for the pages that were also included on Grokipedia.

\begin{figure}[H]
    \centering
    \includegraphics[width=\linewidth]{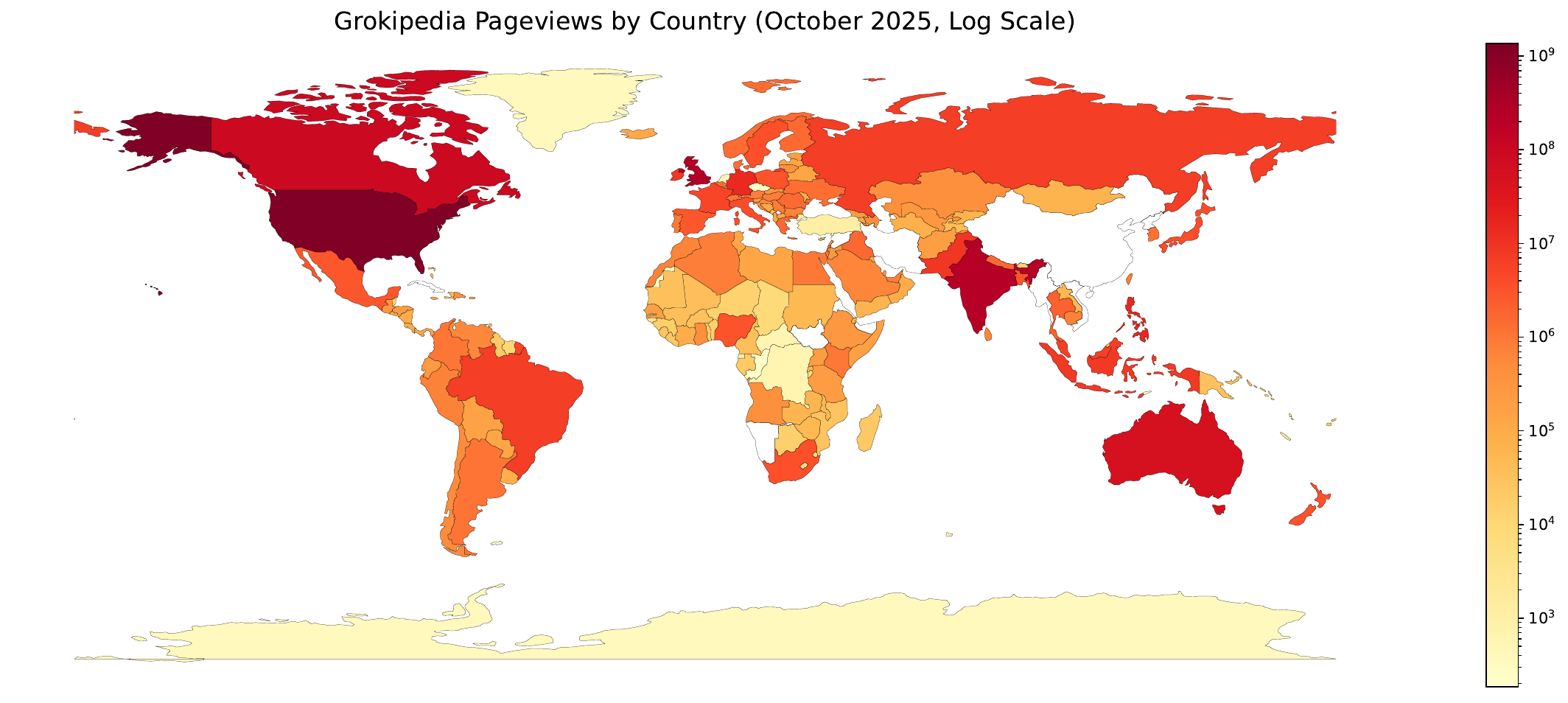}
    \caption{A map of geographic pageview distribution to English Wikipedia from 27 September to 27 October 2025.}
    \label{fig:pageview_map}
\end{figure}

\subsection{Comparing Grokipedia and Wikipedia entries by similarity and article type}
\label{app-subsec:comp}

A comparison of Wikipedia and Grokipedia articles across the spectrum of average chunk similarity, split up by a variety of sub-criteria: the presence or absence of a CC-license, whether the page has been deemed ``Controversial'' by the English Wikipedia community, whether the page is about an elected official.

\subsubsection{With and without a CC-license}%
\begin{table}[H]
    \centering
    \renewcommand{\arraystretch}{1.15}
    \setlength{\tabcolsep}{5pt}
    \begin{tabular}{|p{2cm}|p{1.3cm}|p{1.7cm}|p{5.5cm}|p{5.5cm}|}
        \hline
        \textbf{Entry} & \textbf{Article Type} & \textbf{Similarity} & \textbf{Wikipedia} & \textbf{Grokipedia} \\
        \hline

        \textbf{Sono Sachiko} & With license & High &
        Sachiko's father was Count Sono Motosachi; she was known as Kogiku Tenji. She gave birth to two sons and six daughters, several of whom died prematurely. Her children with Emperor Meiji include the following members of the Japanese imperial family. &
        Sachiko's father was Count Sono Motosachi; she was known as Kogiku Tenji. She gave birth to two sons and six daughters, several of whom died prematurely. Her children with Emperor Meiji include the following members of the Japanese imperial family. \\
        \hline

        \textbf{Mejia Thermal Power Station} & Without license & High &
        Mejia Thermal Power Station is located at Durlabhpur, Bankura, 35 km from Durgapur city in West Bengal. The power plant is one of the coal based power plants of DVC. Commissioned in 1996, MTPS is the largest thermal power plant, in terms of generating capacity, in the state of West Bengal as well as among other DVC power plants. &
        Mejia Thermal Power Station is located at Durlabhpur, Bankura, 35 km from Durgapur city in West Bengal. The power plant is one of the coal based power plants of DVC. Commissioned in 1996, MTPS is the largest thermal power plant, in terms of generating capacity, in the state of West Bengal as well as among other DVC power plants. \\
        \hline

        \textbf{Ellen David} & With license & Low &
        Ellen David began her acting career in 1988, with early roles in television series such as \textit{Night Heat} and \textit{Street Legal}. She has appeared in live-action films including \textit{Mambo Italiano} (2003), \textit{A Walk on the Moon} (1999), and more recently \textit{Brooklyn} (2015), \textit{Goon} (2011), and \textit{Miss Boots} (2024). She is particularly known for her extensive voice work in animated series, voicing characters in \textit{Arthur}, \textit{Caillou}, and \textit{Sagwa, the Chinese Siamese Cat}. &
        Her other animated and live-action roles include: \textit{Tripping the Rift}, \textit{Arthur}, \textit{Mambo Italiano}, \textit{Law \& Order}, \textit{Ciao Bella}, \textit{Mona the Vampire}, \textit{Street Legal}, and many others across film and television. \\
        \hline

        \textbf{Soylent} & Without license & Low &
        Soylent is a brand of meal replacement powders and shakes produced by Soylent Nutrition, Inc., founded in 2013 and headquartered in Los Angeles, California. The name derives from a fictional food in the 1966 novel \textit{Make Room! Make Room!}. &
        Soylent is a brand of plant-based meal replacement products, including ready-to-drink shakes, powders, and bars, formulated to deliver complete nutrition comprising macronutrients, fiber, and essential vitamins and minerals. The products emphasize convenience, sustainability, and affordability. \\
        \hline
    \end{tabular}
    \caption{A comparison of selected articles across the per-article similarity distribution, with license information.}
    \label{app-tab:comp-license}
\end{table}

\subsubsection{Controversial articles}%
\begin{table}[H]
    \centering
    \renewcommand{\arraystretch}{1.15}
    \setlength{\tabcolsep}{5pt}
    \begin{tabular}{|p{2cm}|p{1.7cm}|p{5.5cm}|p{5.5cm}|}
        \hline
        \textbf{Entry} & \textbf{Similarity} & \textbf{Wikipedia} & \textbf{Grokipedia} \\
        \hline

        \textbf{Christopher Paul Neil} & High &
        Christopher Paul Neil (born February 6, 1975), better known as Mr. Swirl Face, is a Canadian teacher who was convicted of child sexual abuse. He was the subject of a highly publicized Interpol investigation of the child sexual abuse of at least 12 young boys in Vietnam, Cambodia, and Thailand, primarily owing to the Internet release of images depicting the abuse. He was arrested by Royal Thai Police in October 2007. &
        Christopher Paul Neil (born February 6, 1975), better known as Mr. Swirl Face, is a Canadian teacher who was convicted of child sexual abuse in Thailand and Canada. He was the subject of a highly publicized Interpol investigation into the child sexual abuse of at least 12 young boys in Vietnam, Cambodia, and Thailand, primarily owing to the Internet release of images depicting the abuse. He was arrested by Royal Thai Police in October 2007 and later arrested in Canada in 2012 for related offenses, leading to additional convictions and his release in 2017. \\
        \hline

        \textbf{Media bias in the United States} & Low &
        The history of media bias in the United States has evolved from overtly partisan newspapers in the 18th and 19th centuries to professional journalism with ethical standards in the 20th and 21st centuries. &
        Media bias in the United States refers to a left-liberal slant in major mainstream outlets, attributed to ideological imbalance in journalism and resulting in disproportionate emphasis on progressive narratives and unfavorable portrayals of conservative viewpoints. \\
        \hline

        \textbf{Masculism} & Low &
        Masculism or masculinism may refer to ideologies and socio-political movements seeking to eliminate discrimination against men or promote masculine ideals. &
        Masculism is a socio-political ideology advocating for men's rights and addressing gender-based disadvantages affecting males, often positioned as a counterpart to feminism. \\
        \hline

    \end{tabular}
    \caption{A comparison of selected articles across the per-article similarity distribution, from the controversial articles subset.}
    \label{app-tab:comp-controversial}
\end{table}

\subsubsection{Elected officials in the US and UK}
\label{app-subsubsec:electeds}

\begin{table}
    \centering
    \makebox[\textwidth][c]{%
    \begin{minipage}{1.2\textwidth} 
        \centering
        \renewcommand{\arraystretch}{1.15}
        \setlength{\tabcolsep}{5pt}
        \begin{tabular}{|p{1.4cm}|p{1.7cm}|p{5.2cm}|p{7.5cm}|}
            \hline
            \textbf{Entry} & \textbf{Similarity} & \textbf{Wikipedia} & \textbf{Grokipedia} \\
            \hline

            \textbf{Chris Deluzio} & High &
            Christopher Raphael Deluzio (born July 13, 1984) is an American politician, attorney, and former U.S. Navy officer serving as the U.S. representative for Pennsylvania's 17th congressional district since 2023. He is a member of the Democratic Party. The district encompasses most of the northwestern suburbs and exurbs of Pittsburgh, and includes the entirety of Beaver County. &
            Christopher Raphael Deluzio (born July 13, 1984) is an American attorney, naval veteran, and politician serving as the U.S. representative for Pennsylvania's 17th congressional district since 2023. A Democrat, he represents a district encompassing Beaver County and portions of Allegheny County in western Pennsylvania. \\
            \hline

            \textbf{Keir Starmer} & Low &
            ...was Director of Public Prosecutions from 2008 to 2013...\par\vspace{21.5mm}
            ...Starmer led Labour to a landslide victory at the 2024 general election...&
            ...was Director of Public Prosecutions from 2008 to 2013, \textbf{overseeing reforms to criminal justice procedures, though his tenure drew criticism for low prosecution rates in child sexual exploitation cases and decisions not to pursue charges in high-profile matters such as the Jimmy Savile complaints.}...\par
            ...Starmer led Labour to a \textbf{parliamentary landslide in the July 2024 general election, securing 412 seats despite a modest 33.7\% vote share—the lowest for a majority government since 1832}... \\
            \hline

            \textbf{Pramila Jayapal} & Low &
            ...serving as the U.S. representative from Washington’s 7th congressional district since 2017... \par\vspace{17mm}
            ...Before entering electoral politics, Jayapal was a Seattle-based civil rights activist, serving until 2012 as the executive director of OneAmerica, an immigrant advocacy group...&
            ...serving as the U.S. Representative for Washington’s 7th congressional district since 2017... \par
            \textbf{Born in Chennai, India, Jayapal immigrated to the United States at age 16 to attend Georgetown University (A.B., 1986), later earning an M.B.A. from Northwestern University in 1990.} \par
            ...founded OneAmerica, an immigrant advocacy organization in Washington state, where she served as executive director for 12 years, focusing on \textbf{public health and human rights issues affecting immigrant communities}... \par
            ...\textbf{She has faced notable controversies, including backlash for describing Israel as a ``racist state'' during a 2023 event (which she later clarified), and for initially questioning a House resolution condemning Hamas’s use of rape as a weapon of war before issuing a statement unequivocally denouncing such acts. These positions reflect her emphasis on human-rights-based foreign policy and divergence from mainstream Democratic views.} \\
            \hline
        \end{tabular}
        \caption{Comparison of Wikipedia and Grokipedia entries for elected officials in the US and UK, with ellipses indicating identical or omitted text.}
        \label{app-tab:comp-elected}
    \end{minipage}%
    }
\end{table}

\clearpage

\subsection{Top 100 domains}

The domain type composition of the top 100 most-cited sources on both Wikipedia and Grokipedia is very different; Grokipedia cites notably more academic sources and UGC.

\begin{figure}[H]
    \centering
    \includegraphics[width=1\linewidth]{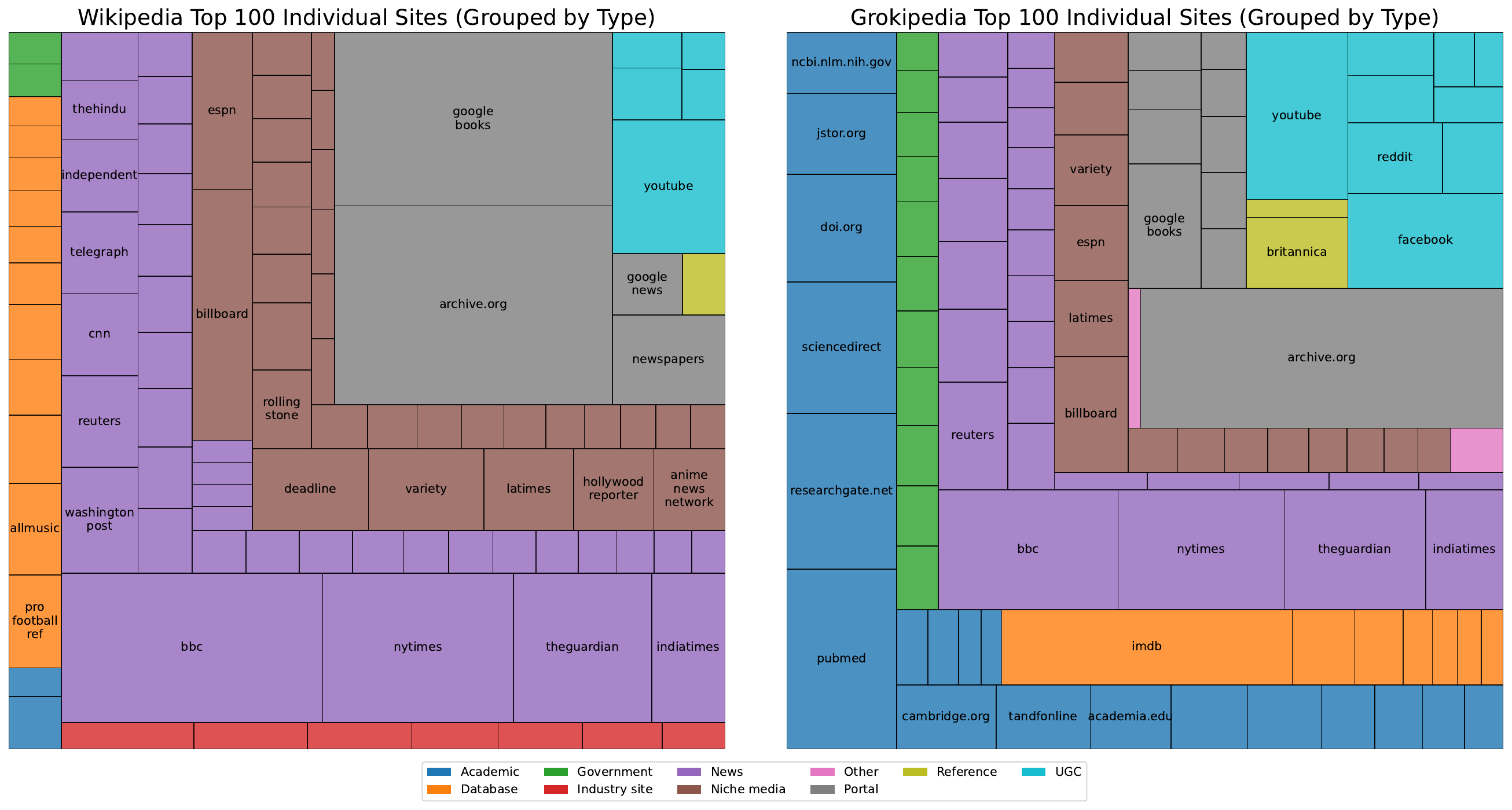}
    \caption{A treemap graph showing the proportion of citations going to each individual domain in the top 100 domains on Wikipedia (left) and Grokipedia (right). Color indicates domain type}
    \label{fig:t100-treemap}
\end{figure}

\subsection{Source reliability}

\begin{figure}[H]
    \centering
    \includegraphics[width=\linewidth]{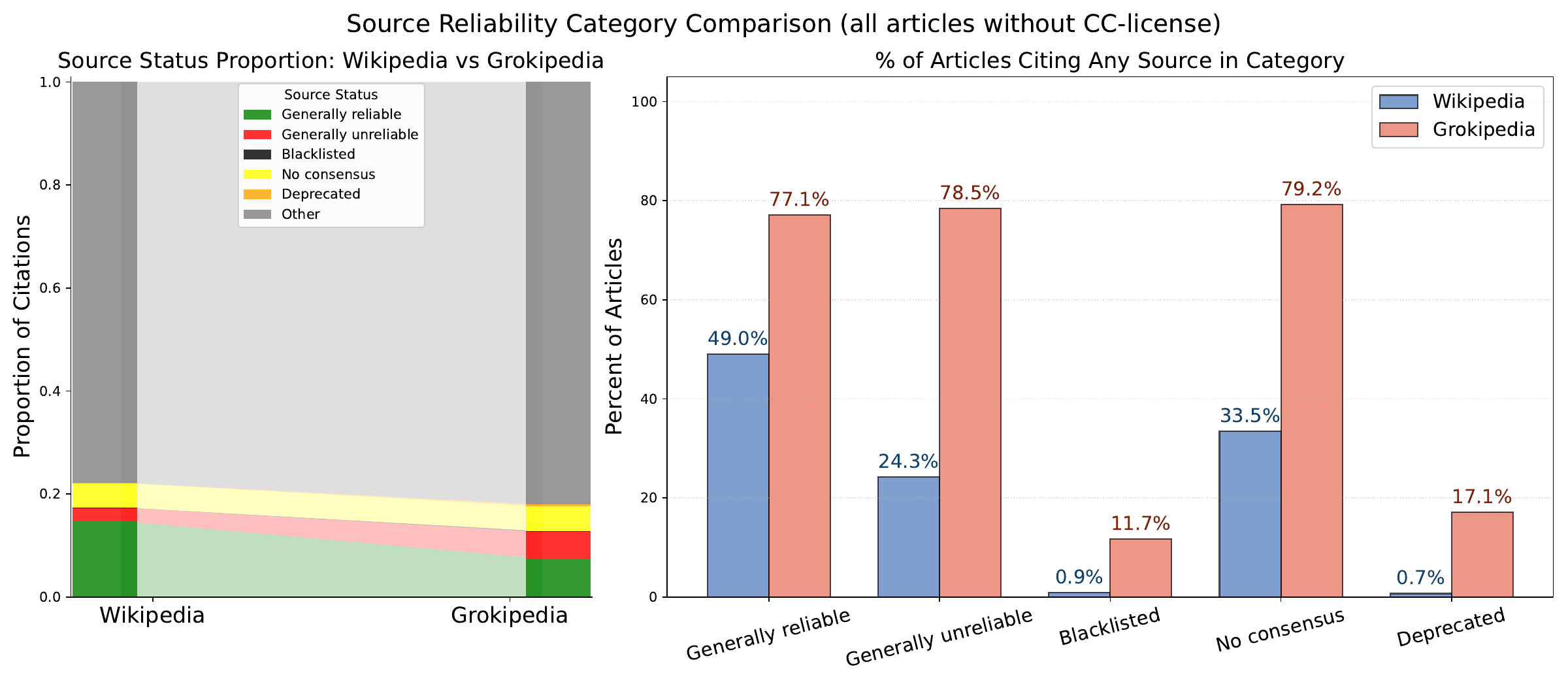}
    \caption{(Left) The relative proportion of Perennial Source list categories for non-CC-licensed Grokipedia articles and their Wikipedia counterparts. (Right) The percentage of articles in that subset that contain at least one of each category of citation in the two corpora.}
    \label{fig:overall-grok-wp-cite-comp-no-cc}
\end{figure}

\begin{figure}[H]
    \centering
    \includegraphics[width=1\linewidth]{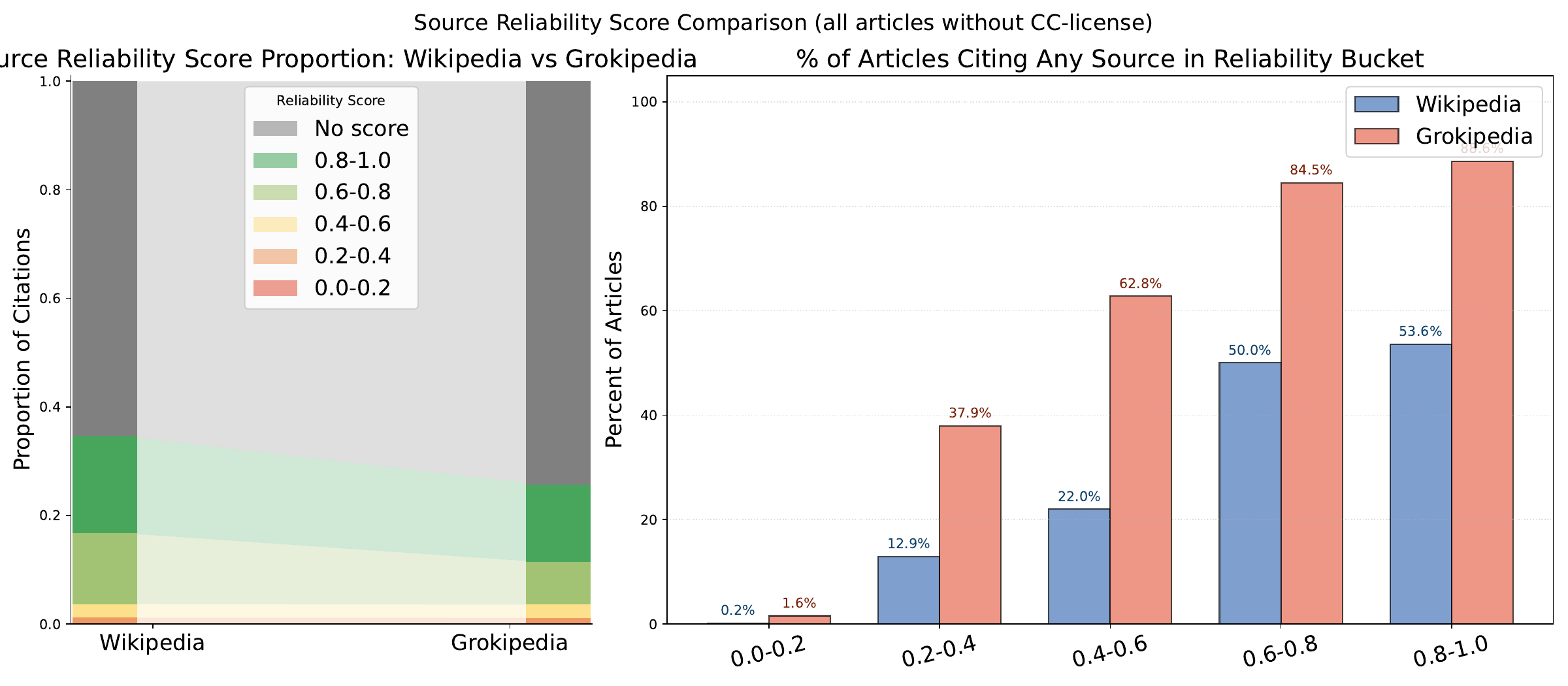}
    \caption{(Left) The relative proportion of Lin et al. score reliability in Wikipedia and Grokipedia citations without a CC license. (Right) The percentage of articles that contain at least one domain in each score bucket in the two corpora.}
    \label{fig:overall-grok-wp-cite-lin-no-cc}
\end{figure}

\subsection{Article topic}

\begin{figure}[H]
    \centering
    \includegraphics[width=\linewidth]{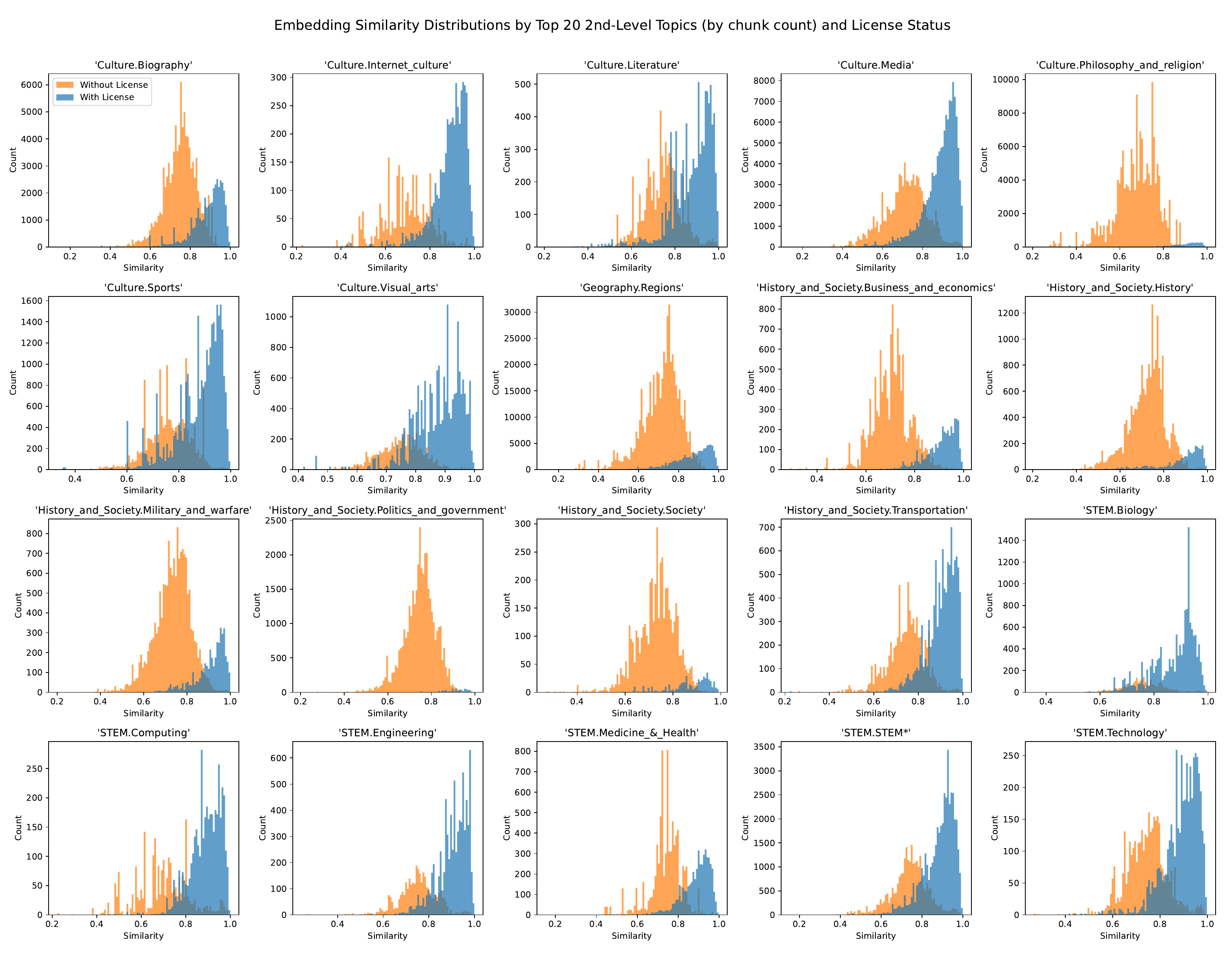}
    \caption{Embedding chunk similarity between Grokipedia and Wikipedia by predicted article subtopic in the random 30,000 article subsample.}
    \label{fig:topic-comp}
\end{figure}